\documentclass[aps,pra,reprint,amsfonts,amssymb,superscriptaddress,floatfix,eprint]{revtex4-2}
\usepackage{graphicx}
\usepackage{amsmath}
\DeclareMathOperator\arctanh{arctanh}
\usepackage{dcolumn}
\usepackage{color}
\usepackage{calc}
\usepackage{MnSymbol}
\usepackage{hyperref}
\usepackage{hypernat}
\usepackage{enumitem}
\usepackage{float}
\usepackage{relsize}
\usepackage{varwidth}
\usepackage{tasks}
\usepackage[utf8]{inputenc}
\usepackage[T1]{fontenc}
\hypersetup{colorlinks=true, citecolor=magenta, urlcolor=blue}

\begin{document}

\title{Arbitrary-angle rotation of the polarization of a dipolar Bose-Einstein condensate}
\author{S. B. Prasad}
\email{srivatsa.badariprasad@unimelb.edu.au}
\affiliation{School of Physics, University of Melbourne, Melbourne, 3010, Australia}
\author{B. C. Mulkerin}
\affiliation{Centre for Quantum and Optical Science, Swinburne University of Technology, Melbourne, 3122, Australia}
\author{A. M. Martin}
\email{martinam@unimelb.edu.au}
\affiliation{School of Physics, University of Melbourne, Melbourne, 3010, Australia}

\date{\today}

\begin{abstract}
We have employed the theory of harmonically trapped dipolar Bose-Einstein condensates to examine the influence of a uniform magnetic field that rotates at an arbitrary angle to its own orientation. This is achieved by semi-analytically solving the dipolar superfluid hydrodynamics of this system within the Thomas-Fermi approximation and by allowing the body frame of the condensate's density profile to be tilted with respect to the symmetry axes of the nonrotating harmonic trap. This additional degree of freedom manifests itself in the presence of previously unknown stationary solution branches for any given dipole tilt angle. We also find that the tilt angle of the stationary state's body frame with respect to the rotation axis is a nontrivial function of the trapping geometry, rotation frequency and dipole tilt angle. For rotation frequencies of at least an order of magnitude higher than the in-plane trapping frequency, the stationary state density profile is almost perfectly equivalent to the profile expected in a time-averaged dipolar potential that effectively vanishes when the dipoles are tilted along the `magic angle', $54.7 \deg$. However, by linearizing the fully time-dependent superfluid hydrodynamics about these stationary states, we find that they are dynamically unstable against the formation of collective modes, which we expect would result in turbulent decay.
\end{abstract}

\maketitle
\section{\label{sec:level0}Introduction}
The last fifteen years in the field of ultracold quantum gases have been marked by a growing prominence of the study of systems with significant two-body dipole-dipole interactions. In 2005, the family of Bose-Einstein condensates (BECs) expanded to include $^{52}$Cr~\cite{prl_94_16_160401_2005, prl_95_15_150406_2005}, the first experimentally realized BEC of atoms that boasted large magnetic dipole moments. Since then, the experimental investigation of dipolar BECs has expanded with the successful condensation of various isotopes of dysprosium~\cite{prl_107_19_190401_2011} and erbium~\cite{prl_108_21_210401_2012}, both of which feature larger magnetic dipole moments than chromium. These systems have been of considerable interest to both the theoretical and experimental communities as they offer a robust, flexible platform for investigating long-range, anisotropic interactions in a many-body quantum system~\cite{physrep_464_3_71-111_2008, repprogphys_72_12_126401_2009, chemrev_112_9_5012-5061_2012, jphyscondesmatter_29_10_103004_2017}. The interplay of these interactions with the short-range interactions ubiquitous in atomic BECs underpins the presence of exotic phenomena that are not present in nondipolar BECs, examples of which include magnetostriction~\cite{nature_448_7154_672-675_2007}, dipolar interaction--induced Landau instabilities~\cite{prl_101_8_080401_2008, natphys_4_3_218-222_2008}, anisotropic superfluidity~\cite{prl_121_3_030401_2018}, the roton mode~\cite{natphys_14_5_442-446_2018, prl_122_18_183401_2019}, and the formation of self-bound ultradilute droplets~\cite{nature_530_7589_194-197_2016, nature_539_7628_259-262_2016, prl_116_21_215301_2016, prx_6_4_041039_2016} that are able to self-organize into a supersolid phase~\cite{prx_9_1_011051_2019, prl_122_13_130405_2019, prx_9_2_021012_2019}.

A major facet of the study of dipolar BECs is the presence of nonzero angular momentum, and the associated existence of quantum vortices, in these systems~\cite{jphyscondesmatter_29_10_103004_2017}. Of particular interest is the structure, dynamics and excitations of vortices and the lattices that they form~\cite{pra_73_6_061602r_2006, prl_100_24_245302_2008, prl_100_9_090406_2008, pra_78_4_043614_2008, pra_79_1_013621_2009, pra_79_21_063622_2009, jphysb_43_13_135301_2010, eurjphysd_61_2_695-699_2011, prl_111_17_170402_2013, jphysb_47_16_165301_2014}, as well as their formation via trap rotation~\cite{prl_98_15_150401_2007, laserphys_18_5_322_2008, pra_80_3_033617_2009}. In the context of rotation, another area of interest has been the possibility of controlling the nature of the dipole-dipole interaction between the constituent atoms in a dipolar BEC by rapidly rotating the dipolar moment polarizations with the aid of an externally applied magnetic field. First proposed in 2002 by Giovanazzi, et al.~\cite{prl_89_13_130401_2002}, this idea was experimentally realized in 2018 by Tang, et al~\cite{prl_120_23_230401_2018}. Specifically, a time-averaged dipolar interaction may be used in lieu of the full time-dependent dipolar interaction induced by the rotating dipoles when the dipole rotation frequency is much greater than any inverse timescale in the system except for the Larmor frequency. Since the time-averaged interaction is dependent on the angle made by the rotating dipole polarization with respect to the rotation axis, experimentalists are able to effectively tune the dipolar interaction strength, and sign, by altering the orientation of the rapidly rotating magnetic field. However, the experimental realization of the rotational tuning scheme reported a significantly diminished condensate lifetime, and subsequent theoretical analyses of dipolar BECs polarized in the plane orthogonal to the rotation axis demonstrated that dynamical instabilities seeded by the rotation may have been a contributing factor to this unexpectedly short lifetime~\cite{prl_122_5_050401_2019, pra_101_4_043606_2020}.

In this article we theoretically address the effect of rotating the polarizing field without explicitly invoking the rapidly rotating, time-averageable limit, and extend our previous work on dipole polarizations rotating in-plane~\cite{prl_122_5_050401_2019, pra_100_2_023625_2019} to polarizations aligned at an arbitrary angle to the rotation axis. To this end, in Sec.~\ref{sec:level1} we introduce a reference frame for conveniently dealing with the tilted rotating polarizing field. We work in the interaction-dominated Thomas-Fermi (TF) regime~\cite{pra_78_4_041601r_2008} and, in Sec.~\ref{sec:level2} we merge the formalisms introduced in distinct studies for analyzing in-plane dipole rotation in dipolar BECs~\cite{prl_122_5_050401_2019, pra_100_2_023625_2019} and tilted rotation of the harmonic confinement of nondipolar BECs~\cite{pra_101_6_063638_2020} to obtain self-consistency relations that completely specify the stationary solutions of this system. In Sec.~\ref{sec:level3a} we find that the semi-analytical solution of these self-consistency relations predicts the existence of previously-unknown stationary solution branches as a consequence of allowing the body frame of the condensate density to be tilted at an arbitrary angle to that of the harmonic trapping. We also find, in Sec.~\ref{sec:level3b}, that for dipole rotation frequencies that are more than an order of magnitude larger than the in-plane trapping frequency, the self-consistency relations reproduce the stationary solution profiles that are satisfied when the rotation frequency is zero and the dipolar interaction is replaced by its time-averaged analog~\cite{pra_71_3_033618_2005, pra_82_3_033612_2010}. Notably, the influence of the dipolar interaction effectively vanishes when the rotating dipole moments are aligned at the so-called `magic angle', $\theta_{\text{c}} = \arccos(1/\sqrt{3}) \approx 54.7 \deg$ to the rotation axis.

However, in Sec.~\ref{sec:level4} we find that the dynamical instabilities that have previously been demonstrated to plague the TF stationary states, when the in-plane--aligned dipole polarization rotates rapidly~\cite{prl_122_5_050401_2019, pra_101_4_043606_2020}, are also present for almost all dipole tilting angles whenever the atomic dipole moment is finite. In the TF regime we also find that the dimensionless chemical potential of the stationary solution predicted by the time-averaged dipolar interaction is at least $10$ times higher than the in-plane trapping frequency. This suggests that it may be necessary for the dipole polarization's rotation frequency to be two or more orders of magnitude larger than the in-plane trapping frequency in order for a substantial number of rotation cycles to occur before the onset of the dynamical instability, which we expect would result in the turbulent decay of the TF state.

\section{\label{sec:level1}Tilted Rotation of the Polarization}
We consider a dilute, scalar BEC of $N$ atoms with dipole-dipole interactions that are mediated by a spatially-uniform magnetic field $\mathbf{B}(t)$ rotating at an angular frequency, $\mathbf{\Omega}$. The atoms are assumed to have a mass $m$, experience an effective contact two-body interaction of strength $g$, and enjoy a magnetic dipole moment $\mu_{\text{d}}$. Provided that $N$ is sufficiently large and that the value of $\varepsilon_{\text{dd}}$ is less than unity, the zero temperature mean-field behavior of this condensate is well described by the Thomas-Fermi--approximated dipolar superfluid hydrodynamic equations~\cite{physrep_464_3_71-111_2008, repprogphys_72_12_126401_2009, chemrev_112_9_5012-5061_2012, jphyscondesmatter_29_10_103004_2017}. In a reference frame co-rotating with the magnetic field, these equations govern the time-dependence of the condensate's number density, $n$, and superfluid velocity field, $\mathbf{v}$, and are given by~\cite{physrep_464_3_71-111_2008, repprogphys_72_12_126401_2009, chemrev_112_9_5012-5061_2012, jphyscondesmatter_29_10_103004_2017, pra_78_4_041601r_2008}:
\begin{gather}
    \frac{\partial n}{\partial t} = -\nabla\cdot\left[n\left(\mathbf{v}-\mathbf{\Omega}\times\mathbf{r}\right)\right], \label{eq:continuity} \\
  m\frac{\partial\mathbf{v}}{\partial t} = -\nabla\left\lbrace\frac{m\mathbf{v}^2}{2} - m\mathbf{v}\cdot(\mathbf{\Omega}\times\mathbf{r}) + V_{\text{T}} + gn + U_{\text{dd}}\otimes n\right\rbrace. \label{eq:euler}
\end{gather}
Here the symbol $\otimes$ denotes the convolution of $n(\mathbf{r}, t)$ with the dipolar interaction kernel, $U_{\text{dd}}(\mathbf{r}, t)$, which is defined as~\cite{pra_61_5_051601r_2000, prl_85_9_1791-1794_2000, pra_63_5_053607_2001}
\begin{equation}
    U_{\text{dd}}(\mathbf{r}, t) = \frac{C_{\text{dd}}}{4\pi}\frac{1-3\left[\widehat{\mathbf{B}}(t)\cdot\hat{\mathbf{r}}\right]^2}{r^3} = \frac{\mu_0\mu_{\text{d}}^2}{4\pi}\frac{1-3\left[\widehat{\mathbf{B}}(t)\cdot\hat{\mathbf{r}}\right]^2}{r^3}, \label{eq:DDIkernel}
\end{equation}
with $\mu_0$ the vacuum permeability. The density of the BEC obeys the normalization condition
\begin{equation}
  \int d^3r\,n(\mathbf{r}, t) = N. \label{eq:normalization}
\end{equation}
In Eq.~\eqref{eq:euler} the true short-ranged two-body interaction is replaced by an effective contact interaction with a strength $g$ is defined in terms of $a_{\text{s}}$, the $s$-wave scattering length of the full interaction potential, as~\cite{pra_63_5_053607_2001, ronen_bortolotti_2006}
\begin{equation}
    g = \frac{4\pi\hbar^2a_{\text{s}}}{m}. \label{eq:contactint}
\end{equation}
In addition, the relative dipolar interaction ratio~\cite{repprogphys_72_12_126401_2009},
\begin{equation}
    \varepsilon_{\text{dd}} = \frac{C_{\text{dd}}}{3g}, \label{eq:edddef}
\end{equation}
is used to parametrize the dipolar interaction. The external confinement of the condensate enters Eq.~\eqref{eq:euler} through the trapping term, $V_{\text{T}}$, which we take to be a time-independent, cylindrically symmetric harmonic trap:
\begin{equation}
    V_{\text{T}}(\mathbf{r}) = \frac{m\omega_{\perp}^2}{2}\left(\rho^2 + \gamma^2z^2\right). \label{eq:trapuprightcoord}
\end{equation}
We also define a harmonic trapping length, $l_{\perp}$, as
\begin{equation}
    l_{\perp} = \sqrt{\frac{\hbar}{m\omega_{\perp}}}. \label{eq:harmoniclength}
\end{equation}
The superfluid hydrodynamic equations constitute a reformulation of the dipolar Gross-Pitaevskii equation (dGPE) in which the condensate order parameter, $\psi$, has been replaced by the density and superfluid velocity. Together with the condensate phase, $S$, these are related to $\psi$ via the Madelung transformation~\cite{pitaevskiistringaribec, pethicksmithbecdilutegases}:
\begin{align}
  \psi &= \sqrt{n}e^{iS}, \label{eq:nsdef} \\
  \mathbf{v} &= \frac{\hbar\nabla S}{m}. \label{eq:vdef}
\end{align}

We also note that Eq.~\eqref{eq:euler} contains two notable approximations that are deemed to be appropriate in the context of this work. When $Na_{\text{s}}(1-\varepsilon_{\text{dd}}) \gg l_{\perp}$ an additional quantum pressure term on the right-hand side of Eq.~\eqref{eq:euler}, $\nabla\left[\hbar^2\nabla^2(\sqrt{n})/(2m\sqrt{n})\right]$, is negligible due to the minimal effects of zero-point kinetic energy fluctuations in the condensate~\cite{pra_51_2_1382-1386_1995, prl_76_1_6-9_1996, pra_78_4_041601r_2008}. For the remainder of this work we work in the TF regime and thus neglect the zero-point fluctuations. While the application of Eqs.~\eqref{eq:continuity} and \eqref{eq:euler} to trapped, dilute, dipolar BECs has predicted several properties such as an anisotropic Bogoliubov spectrum~\cite{pra_61_5_051601r_2000, prl_85_9_1791-1794_2000, prl_121_3_030401_2018}, the roton mode~\cite{prl_90_25_250403_2003, prl_98_3_030406_2007, natphys_14_5_442-446_2018, prl_122_18_183401_2019}, and exotic vortex behavior~\cite{pra_73_6_061602r_2006, prl_100_24_245302_2008, prl_111_17_170402_2013, pra_100_2_023625_2019}, the experimental discovery of self-bound quantum droplets~\cite{nature_530_7589_194-197_2016, nature_539_7628_259-262_2016, prl_116_21_215301_2016, prx_6_4_041039_2016} and the realization of supersolidity~\cite{prx_9_1_011051_2019, prl_122_13_130405_2019, prx_9_2_021012_2019} in recent years has necessitated the extension of this theory to account for the contributions of fluctuations of $\psi$ beyond the mean field~\cite{pra_86_6_063609_2012, pra_94_3_033619_2016}. Provided that $\varepsilon_{\text{dd}} \lesssim 1$, the effect of these fluctuations is generally insignificant and thus we expect that Eqs.~\eqref{eq:continuity} and \eqref{eq:euler} provide robust theoretical predictions in this regime.

The vast majority of experimental and theoretical studies of dipolar BECs have assumed a dipole orientation along one of the harmonic trap's principal axes, $\hat{x}$, $\hat{y}$, or $\hat{z}$. Indeed, with regards to the rotation of either the trapping or the dipole orientation, we are not aware of any previous studies in the TF regime that systematically investigated dipole alignments that were not either parallel or orthogonal to the angular frequency vector, $\Omega$. These assumptions greatly simplify the subsequent theoretical analysis due to the resulting symmetries that constrain the solutions of Eqs.~\eqref{eq:continuity} and \eqref{eq:euler}. However, in order to provide a full treatment of the time-averaging property of the dipolar interaction via the hydrodynamic equations, it is necessary to account for dipole alignments that are at an arbitrary angle to the rotation axis and to any one of the trap's principal axes. To this end we define $\theta$ as the angle made by the dipole-polarizing magnetic field with respect to its rotation axis, which we take to be $\hat{z}$ without loss of generality. We also assume that the rotation frequency is constant, i.e. $\mathbf{\Omega} = \Omega\hat{z}$. Defining the $x$-$y$ axes such that the projection of the magnetic field upon the $y$-axis is zero at $t = 0$, we have
\begin{equation}
    \mathbf{B}(t) = B\left\lbrace\sin\theta[\cos(\Omega t)\hat{x} + \sin(\Omega t)\hat{y}] + \cos\theta\hat{z}\right\rbrace, \label{eq:rotbfield}
\end{equation}
the resulting dipolar interaction kernel is given by
\begin{equation}
    \frac{U_{\text{dd}}(\mathbf{r}, t)}{C_{\text{dd}}} = \frac{r^2-3\left\lbrace\sin\theta[\cos(\Omega t)x + \sin(\Omega t)y] + \cos\theta z\right\rbrace^2}{4\pi r^5}. \label{eq:DDIkerneltimedep}
\end{equation}
The time-average of this time-dependent kernel, over one full rotation cycle of the magnetic field, is equivalent to~\cite{prl_89_13_130401_2002}
\begin{align}
    \llangle U_{\text{dd}}(\mathbf{r})\rrangle &\equiv \frac{\Omega}{2\pi}\int_0^{\frac{2\pi}{\Omega}}dt\,U_{\text{dd}}(\mathbf{r}, t) \nonumber \\
    &= C_{\text{dd}}\left(\frac{3\cos^2\theta-1}{2}\right)\frac{r^2-3z^2}{4\pi r^5}. \label{eq:DDIkerneltimeavg}
\end{align}

Remarkably the time-averaged dipole-dipole interaction with the rotating magnetic field is equivalent to a dipole-dipole interaction generated by a non-rotating magnetic field aligned along the rotation axis, which we have taken to be the $z$-axis, but with a $\theta$-dependent multiplicative factor, $(3\cos^2\theta-1)/2$. This factor is equal to $1$ when $\theta = 0$ and decreases monotonically until it equals $-1/2$ when $\theta = \pi/2$. The crossover of the time-averaged dipolar interaction strength from positive to negative values indicates that the two-body interaction between head-to-tail--aligned dipole moments undergoes a crossover from attraction to repulsion. The point of crossover occurs when $\theta$ equals a `magic angle'~\cite{prl_89_13_130401_2002},
\begin{equation}
    \theta_{\text{c}} = \arccos(1/\sqrt{3}) \approx 54.7 \deg, \label{eq:magicangle}
\end{equation}
where the time-averaged dipolar interaction vanishes for all values of $C_{\text{dd}}$. Such an angle is well-known in the literature as the `magic angle' for rapid rotation in magnetic resonance studies where unwanted dipolar resonances may be effectively eliminated~\cite{philostransrsoca_299_1452_505-520_1981}. Following from this result it was predicted that for sufficiently high rotation frequencies, albeit lower than the Larmor frequency associated with the magnetic field strength, the dipolar BEC subject to this rotating magnetic field would effectively experience the time-averaged dipolar potential, thereby effectively providing experimentalists a tuning knob for the dipolar potential's strength and sign in the form of the field tilting angle, $\theta$~\cite{prl_89_13_130401_2002}. This magic angle is the same angle made by the dipole polarization and the $z$-axis, such that the dipolar interaction vanishes entirely, for quasi-$1$D dipolar systems whose dynamics are effectively restricted to the $z$-axis~\cite{chemrev_112_9_5012-5061_2012}. However, it is crucial to note that in inherently $3$D dipolar systems, it is only the time-averaged dipolar interaction that vanishes at this magic angle and not the true dipolar interaction.

This theoretical prediction has been the focus of experimental studies that find that for $\Omega \gg \omega_{\perp}$, a dipolar BEC indeed resembles one that experiences the time-averaged dipolar interaction in Eq.~\eqref{eq:DDIkerneltimeavg}, though the condensate lifetime was found to be orders of magnitude lower than expected~\cite{prl_120_23_230401_2018}. Subsequent theoretical studies addressing this unexpectedly short condensate lifetime proposed that a dynamical instability associated with the small, but nonetheless finite, time-dependent effects at high rotation frequency were responsible for an exponential deviation from the ground state of the Thomas-Fermi hydrodynamic equations, Eqs.~\eqref{eq:continuity} and \eqref{eq:euler} over long timescales~\cite{prl_122_5_050401_2019, pra_101_4_043606_2020}. However, while it has been possible to simulate this scenario via direct solution of the dGPE for a magnetic field aligned at any given value of $\theta$~\cite{pra_101_4_043606_2020}, the complementary semi-analytical methods are only applicable when the magnetic field is aligned along one of the trap's principal axes at $t = 0$, i.e. $\theta = 0$ or $\theta = \pi$~\cite{pra_80_3_033617_2009, prl_122_5_050401_2019}.

To achieve this generalization, we define a second co-rotating reference frame, with coordinates $\mathbf{R}$, in which $\mathbf{B} \equiv B\widehat{Z}$. The $\mathbb{SO}(3)$ rotation that achieves this is of the form
\begin{align}
\begin{pmatrix}
X \\
Y \\
Z
\end{pmatrix} =
\begin{pmatrix}
\cos\theta & 0 & -\sin\theta \\
0 & 1 & 0 \\
\sin\theta & 0 & \cos\theta
\end{pmatrix}
\begin{pmatrix}
x \\
y \\
z
\end{pmatrix}. \label{eq:fieldtiltframe}
\end{align}
However, it is not possible to solve Eqs.~\eqref{eq:continuity} and \eqref{eq:euler} immediately using these coordinates because the vector $\mathbf{\Omega}\times\mathbf{r}$ lies in the $x$-$y$ plane whereas the magnetic field lies along the $Z$-axis. The resulting competition between the Coriolis and centrifugal effects associated with the rotating frame and the magnetostriction associated with the dipolar interaction suggests that the principal axes of solutions, whether time-independent or -dependent, are generally unlikely to be coincident with either those of $\mathbf{r}$ or $\mathbf{R}$. Thus we introduce a third co-rotating reference frame, with the Cartesian coordinates $\tilde{\mathbf{r}}$, and relate this new coordinate system to $\mathbf{r}$ and $\mathbf{R}$ via another $\mathbb{SO}(3)$ rotation angle, $\xi$, and the associated transformation:
\begin{align}
\begin{pmatrix}
\tilde{x} \\
\tilde{y} \\
\tilde{z}
\end{pmatrix} &=
\begin{pmatrix}
\cos\xi & 0 & \sin\xi \\
0 & 1 & 0 \\
-\sin\xi & 0 & \cos\xi
\end{pmatrix}
\begin{pmatrix}
X \\
Y \\
Z
\end{pmatrix} \nonumber \\
&=
\begin{pmatrix}
\cos(\theta-\xi) & 0 & -\sin(\theta-\xi) \\
0 & 1 & 0 \\
\sin(\theta-\xi) & 0 & \cos(\theta-\xi)
\end{pmatrix}
\begin{pmatrix}
x \\
y \\
z
\end{pmatrix}. \label{eq:dipoleadjusttiltframe}
\end{align}

\begin{figure}[htbp]
\centering
\includegraphics[width=\linewidth]{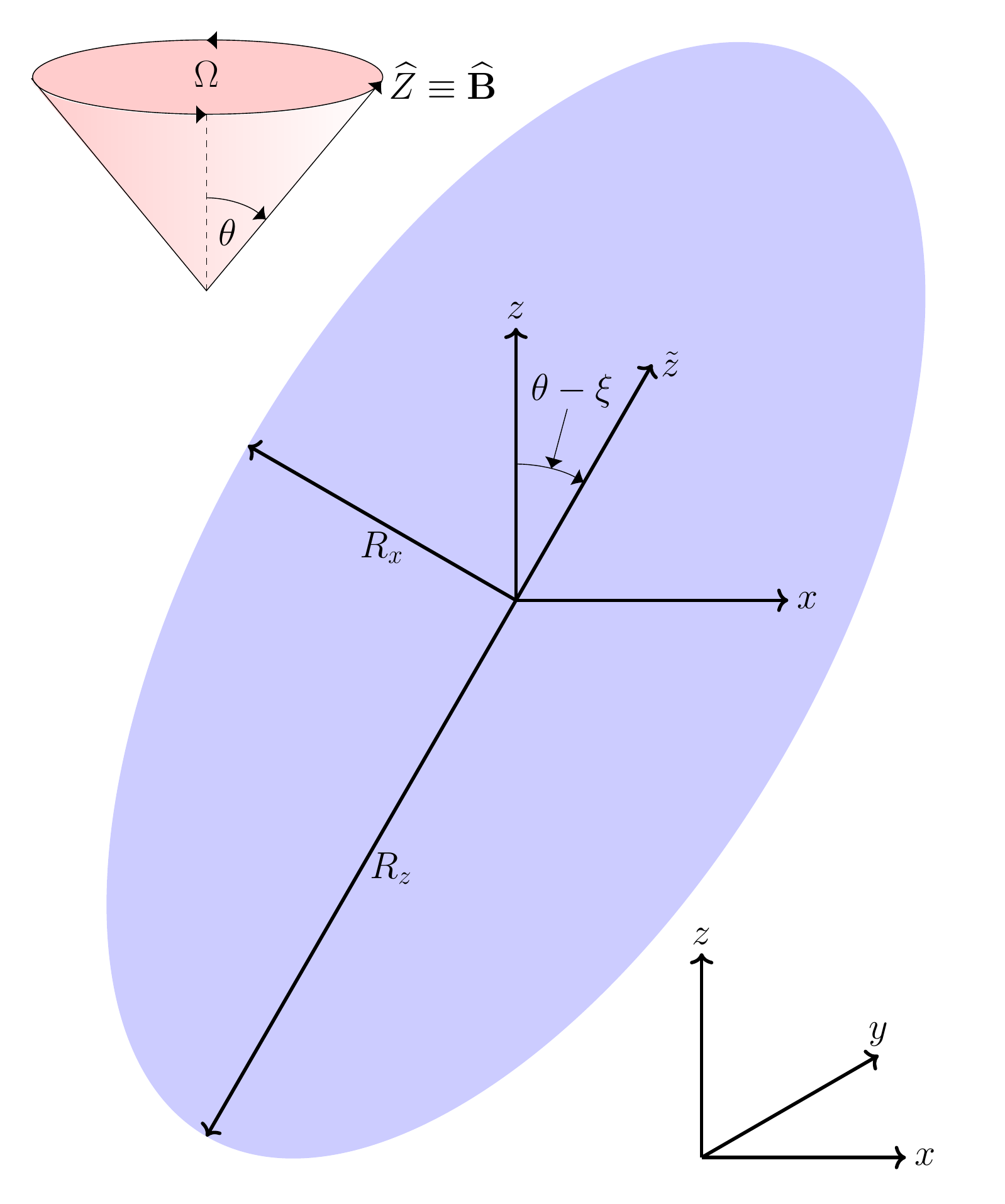}
\caption{Shaded cross-section, at $y = Y = \tilde{y} = 0$, of the ellipsoidal surface of constant density for a Thomas-Fermi stationary state with its semi-axes along the $\tilde{x}$- and $\tilde{z}$-axes, $R_x$ and $R_z$, respectively, illustrated for reference. The Cartesian axes corresponding to the coordinate frames $\tilde{\mathbf{r}}$ and $\mathbf{r}$ are also depicted. Adjacent to the cross-section, a magnetic field cone illustrates the rotation of the polarization with angular frequency, $\Omega$, and tilt angle, $\theta$.}
\label{referenceframes}
\end{figure}

In the coordinate frame $\tilde{\mathbf{r}}$, the trapping potential is given by
\begin{align}
  V_{\text{T}}\left(\tilde{\mathbf{r}}\right) &= \frac{m\omega_{\perp}^2}{2}\left\lbrace\left[\tilde{x}\cos(\theta-\xi) + \tilde{z}\sin(\theta-\xi)\right]^2 + \tilde{y}^2\right\rbrace \nonumber \\
  &+ \frac{m\gamma^2\omega_{\perp}^2}{2}\left[\tilde{x}\sin(\theta-\xi) - \tilde{z}\cos(\theta-\xi)\right]^2, \label{eq:traptilted}
\end{align}
while the effects of the rotation are represented by
\begin{equation}
\mathbf{\Omega}\times\tilde{\mathbf{r}} = \Omega\left[\cos(\theta-\xi)\left(-\tilde{y}\hat{\tilde{x}} + \tilde{x}\hat{\tilde{y}}\right) + \sin(\theta-\xi)\left( \tilde{z}\hat{\tilde{y}} - \tilde{y}\hat{\tilde{z}}\right)\right]. \label{eq:galileitilted}
\end{equation}
Furthermore the dipole polarization in these new coordinates is specified by
\begin{equation}
\mathbf{B} = B\left[\hat{\tilde{x}}\sin\xi + \hat{\tilde{z}}\cos\xi\right]. \label{eq:fieldtilted}
\end{equation}
To clarify the relationship between the co-rotating reference frames, we plot the coordinate axes of $\mathbf{r}$ and $\tilde{\mathbf{r}}$ at constant $y = Y = \tilde{y} = 0$ along with a typical cross-section of an TF surface of constant density in Fig. \ref{referenceframes}. Figure \ref{referenceframes} also contains an illustration of a magnetic field cone depicting the tilted rotation of the polarization vector, which is parallel to the corotating unit vector $\widehat{Z}$.

\section{\label{sec:level2}Thomas-Fermi Stationary Solutions}
To understand the behavior of a dipolar BEC in the TF regime with tilted, rotating dipole moments, we solve for the stationary solutions of Eqs.~\eqref{eq:continuity} and \eqref{eq:euler}. Subsequently, by obtaining the linearized spectrum of collective modes of the dipolar BEC about these stationary states, we can characterize the dynamical stability of the condensate in the TF regime. The stationary solutions of Eqs.~\eqref{eq:continuity} and \eqref{eq:euler} are defined indirectly as the density and velocity profiles that, via Eqs.~\eqref{eq:nsdef} and \eqref{eq:vdef}, correspond to the order parameter $\psi_0(\tilde{\mathbf{r}})$, which satisfies~\cite{pitaevskiistringaribec}
\begin{equation}
  \psi(\tilde{\mathbf{r}}, t) = \psi(\tilde{\mathbf{r}}, t = 0)\exp(-i\mu t). \label{eq:chempot}
\end{equation}
Here, $\mu$ is the condensate's chemical potential. Therefore, the stationary state density, $n_{\text{TF}}$, and velocity, $\mathbf{v}_{\text{TF}}$, obey
\begin{gather}
  0 = \nabla\cdot\left[n_{\text{TF}}\left(\mathbf{v}_{\text{TF}}-\mathbf{\Omega}\times\tilde{\mathbf{r}}\right)\right], \label{eq:contstat} \\
  \mu = \frac{m\mathbf{v}_{\text{TF}}^2}{2} - m\mathbf{v}_{\text{TF}}\cdot(\mathbf{\Omega}\times\tilde{\mathbf{r}}) + V_{\text{T}} + gn_{\text{TF}} + U_{\text{dd}}\otimes n_{\text{TF}}. \label{eq:eulerstat}
\end{gather}

In line with the stationary solutions for the analogous nondipolar system of a nondipolar BEC in an rotating, tilted harmonic trap, we find that the solutions of Eqs. ~\eqref{eq:contstat} and \eqref{eq:eulerstat} are given by the TF density, $n_{\text{TF}}(\tilde{\mathbf{r}})$, and the quadrupolar phase profile, $\mathbf{v}_{\text{TF}}(\tilde{\mathbf{r}})$, which are given by~\cite{pra_101_6_063638_2020}
\begin{gather}
  n_{\text{TF}}(\tilde{\mathbf{r}}) = n_0\left(1 - \mathlarger{\sum_{i\in{x,y,z}}}\frac{\tilde{r}_i^2}{R_i^2}\right)\Theta\left(1 - \mathlarger{\sum_{i\in{x,y,z}}}\frac{\tilde{r}_i^2}{R_i^2}\right), \label{eq:nstat} \\
  \mathbf{v}_{\text{TF}}(\tilde{\mathbf{r}}) = \nabla\left[\alpha\tilde{x}\tilde{y} + \delta\tilde{y}\tilde{z}\right]. \label{eq:vstat}
\end{gather}
Here, $n_0 = 15N/(8\pi R_xR_yR_z)$ is a normalization parameter that ensures that $n_{\text{TF}}$ obeys Eq.~\eqref{eq:normalization}~\cite{pra_51_2_1382-1386_1995, prl_76_1_6-9_1996}. The form of Eq.~\eqref{eq:nstat} shows that the angle $\xi$ in the coordinate transformation given by Eq.~\eqref{eq:dipoleadjusttiltframe} is fixed by the requirement that the principal axes of the TF stationary state density coincide with the Cartesian axes of the $\tilde{\mathbf{r}}$ coordinate frame. The parameters $\lbrace R_i\rbrace$ thus denote the semi-axes of the paraboloid TF profile along the $\tilde{r}_i$-axis. We illustrate these features in the TF density cross-section in Fig.~\ref{referenceframes} by labeling the semi-axes along $\hat{\tilde{x}}$ and $\hat{\tilde{z}}$ as $R_x$ and $R_z$, respectively. Equation~\eqref{eq:vstat} is a generalization of the quadrupolar TF stationary state velocity profile $\mathbf{v}_{\text{TF}} = \alpha\nabla(xy)$ nondipolar BECs, as well as dipolar BECs with the dipole tilt angle $\theta = n\pi/2\,:\,n\in\mathbb{Z}$, in an untilted harmonic trap~\cite{prl_86_3_377-380_2001, prl_98_15_150401_2007, laserphys_18_5_322_2008, pra_80_3_033617_2009, prl_122_5_050401_2019, pra_100_2_023625_2019}. The existence of the additional term $\delta\nabla(\tilde{y}\tilde{z})$ is necessary in order to correctly account for the possibility of nontrivial values of $\theta - \xi$. In other words, together, the nontrivial angle $\xi$ and the phase term $\delta\tilde{y}\tilde{z}$ reconcile the interplay of the rotation, dipolar interaction and trapping in a manner that allows us to employ the conventional TF solution for the harmonic oscillator potential as a stationary state of Eqs.~\eqref{eq:contstat} and \eqref{eq:eulerstat}. However, it is not necessary to include a term proportional to $\nabla(\tilde{z}\tilde{x})$ since the projection of $\mathbf{\Omega}$ upon the $\tilde{y}$-axis is zero~\cite{pra_101_6_063638_2020}. By substituting Eq.~\eqref{eq:nstat} and~\eqref{eq:vstat} into Eq.~\eqref{eq:contstat} and equating the coefficients of the spatial coordinates, we derive the relations
\begin{align}
\alpha &= \left(\frac{R_x^2 - R_y^2}{R_x^2 + R_y^2}\right)\Omega\cos(\theta-\xi), \label{eq:alphadefn} \\
\delta &= \left(\frac{R_z^2 - R_y^2}{R_z^2 + R_y^2}\right)\Omega\sin(\theta-\xi). \label{eq:deltadefn}
\end{align}
Equations~\eqref{eq:alphadefn} and \eqref{eq:deltadefn} are equivalent to the definitions for the respective velocity amplitudes of the stationary states of a nondipolar BEC, in the TF regime, that is trapped by a tilted, rotating harmonic trap~\cite{pra_101_6_063638_2020}. This quadrupolar condensate phase profile also coincides with the classical velocity potential of an inviscid fluid inside an ellipsoidal container rotating about a nonprincipal axis~\cite{lambhydrodynamics, landaulifshitzvol6fluidmechanics, prl_86_3_377-380_2001}; the dipolar BEC in a tilted, rotating frame represents a dipolar, quantum analog of the classical hydrodynamic system. We also note that Eqs.~\eqref{eq:alphadefn} and \eqref{eq:deltadefn} are formally similar to the equations of motion that appear in the context of the rotational energy levels in the tilted-axis cranked-shell model of rotating triaxial nuclei~\cite{prc_65_5_054304_2002}.

The dipolar contribution to Eq.~\eqref{eq:euler} may be evaluated by recasting it in the form of a Coulomb potential, $\phi_{\text{dd}}(\tilde{\mathbf{r}}, t)$, as follows~\cite{prl_92_25_250401_2004, pra_71_3_033618_2005}:
\begin{align}
U_{\text{dd}}(\tilde{\mathbf{r}})\otimes n(\tilde{\mathbf{r}},t)
&= -g\varepsilon_{\text{dd}}\left[n(\tilde{\mathbf{r}},t) + 3\left(\widehat{\mathbf{B}}\cdot\nabla\right)^2\phi_{\text{dd}}(\tilde{\mathbf{r}},t)\right], \label{eq:ddirecast} \\
\phi_{\text{dd}}(\tilde{\mathbf{r}},t) &= \frac{1}{4\pi}\int d^3\tilde{r}'\,\frac{n(\tilde{\mathbf{r}}',t)}{|\mathbf{r}-\mathbf{r}'|}. \label{eq:ddipseudopot}
\end{align}
The potential, $\phi_{\text{dd}}$, obeys Poisson's equation, $\nabla^2\phi_{\text{dd}} = -4\pi n$, allowing for the self-consistent solution of the density via methods developed for the evaluation of the gravitational potential inside self-gravitating ellipsoidal bodies. For the TF density given by Eq.~\eqref{eq:nstat}, these methods yield the exact form of $\phi_{\text{dd}}$
~\cite{prl_92_25_250401_2004, pra_71_3_033618_2005}:
\begin{align}
\phi_{\text{dd}}(\mathbf{r}) &= \frac{n_0\kappa_x\kappa_y}{4}\left(\frac{R_z^2\beta_{000}}{2} - x^2\beta_{100}-y^2\beta_{010}-z^2\beta_{001}\right) \nonumber \\
&+ \frac{n_0\kappa_x\kappa_y}{8R_z^2}(x^4\beta_{200} + y^4\beta_{020}+z^4\beta_{002}) \nonumber \\
&+ \frac{n_0\kappa_x\kappa_y}{4R_z^2}(x^2y^2\beta_{110}+y^2z^2\beta_{011}+x^2z^2\beta_{101}). \label{eq:pseudopotbeta}
\end{align}
Here, $\beta_{ijk}\left(\kappa_x,\kappa_y\right)$ denotes the integral~\cite{pra_71_3_033618_2005, prl_98_15_150401_2007}
\begin{equation}
\beta_{ijk}\left(\kappa_x,\kappa_y\right) = \int_0^{\infty}\frac{\mathrm{d}s}{(\kappa_x^2+s)^{i+\frac{1}{2}}(\kappa_y^2+s)^{j+\frac{1}{2}}(1+s)^{k+\frac{1}{2}}}, \label{eq:beta}
\end{equation}
where $\kappa_x = R_x/R_z$ and $\kappa_y = R_y/R_z$. The nonlocal contribution from Eq.~\eqref{eq:ddirecast} to Eq.~\eqref{eq:eulerstat} is therefore given by
\begin{widetext}
\begin{align}
-\left(\widehat{\mathbf{B}}\cdot\nabla\right)^2\phi_{\text{dd}}\left(\tilde{\mathbf{r}}\right) &= \frac{n_0\kappa_x\kappa_y\left(\beta_{001}\cos^2\xi + \beta_{100}\sin^2\xi\right)}{2} - \frac{n_0\kappa_x\kappa_y\beta_{101}\sin(2\xi)\tilde{x}\tilde{z}}{R_z^2} \nonumber \\
&- \frac{n_0\kappa_x\kappa_y\left[(\beta_{101}\cos^2\xi + 3\beta_{200}\sin^2\xi)\tilde{x}^2 + (\beta_{011}\cos^2\xi + \beta_{110}\sin^2\xi)\tilde{y}^2 + (3\beta_{002}\cos^2\xi + \beta_{101}\sin^2\xi)\tilde{z}^2\right]}{2R_z^2}. \label{eq:ddinttiltanswer}
\end{align}
\end{widetext}

Combining all of the various contributions to Eqs.~\eqref{eq:contstat} and \eqref{eq:eulerstat} allows us to obtain five self-consistency relations for $\lbrace \kappa_x, \kappa_y, \alpha, \delta, \xi\rbrace$. A solution set for these parameters may then be used to find $R_z$ and $\mu$, thereby fully specifying the state of the condensate in the TF regime. These self-consistency relations are obtained by substituting the trial solutions for the density and velocity profiles, Eqs.~\eqref{eq:nstat} and \eqref{eq:vstat}, into Eqs.~\eqref{eq:contstat} and \eqref{eq:eulerstat}, and equating the coefficients of like terms. To this end it is useful to define the effective harmonic trapping frequencies, $\tilde{\omega}_i^2$:
\begin{align}
\tilde{\omega}_x^2 &= \cos^2(\theta-\xi) + \gamma^2\sin^2(\theta-\xi) + \tilde{\alpha}^2 - 2\widetilde{\Omega}\tilde{\alpha}\cos(\theta - \xi), \label{eq:omegaxeff} \\
\tilde{\omega}_y^2 &= 1 + \tilde{\alpha}^2 + \tilde{\delta}^2 + 2\widetilde{\Omega}[\tilde{\alpha}\cos(\theta-\xi)+\tilde{\delta}\sin(\theta-\xi)], \label{eq:omegayeff} \\
\tilde{\omega}_z^2 &= \gamma^2\cos^2(\theta-\xi) + \sin^2(\theta-\xi) + \tilde{\delta}^2 - 2\widetilde{\Omega}\tilde{\delta}\sin(\theta-\xi), \label{eq:omegazeff} \\
S_{xz} &= (1-\gamma^2)\sin\left[2(\theta-\xi)\right] + 2\tilde{\alpha}\tilde{\delta}  \nonumber \\
&- 2\widetilde{\Omega}\left[\tilde{\alpha}\sin(\theta-\xi) + \tilde{\delta}\cos(\theta-\xi)\right]. \label{eq:sxzeff}
\end{align}
Here we have written $\alpha = \tilde{\alpha}\omega_{\perp}$, $\delta = \tilde{\delta}\omega_{\perp}$, and $\Omega = \widetilde{\Omega}\omega_{\perp}$ for the sake of brevity. We also define the following quantities, each involving the $\beta_{ijk}$ integrals, that will allow us to succinctly account for the dipolar contributions to Eq.~\eqref{eq:eulerstat}:
\begin{align}
    \zeta_x &= 1 + \varepsilon_{\text{dd}}\left[\frac{3}{2}\kappa_x^3\kappa_y(3\sin^2\xi\beta_{200} + \cos^2\xi\beta_{101}) - 1\right], \label{eq:zetax} \\
    \zeta_y &= 1 + \varepsilon_{\text{dd}}\left[\frac{3}{2}\kappa_x\kappa_y^3(\sin^2\xi\beta_{110} + \cos^2\xi\beta_{011}) - 1\right], \label{eq:zetay} \\
    \zeta_z &= 1 + \varepsilon_{\text{dd}}\left[\frac{3}{2}\kappa_x\kappa_y(\sin^2\xi\beta_{101} + 3\cos^2\xi\beta_{002}) - 1\right], \label{eq:zetaz} \\
    \chi &= 1+\varepsilon_{\text{dd}}\left[\frac{3}{2}\kappa_x\kappa_y(\sin^2\xi\beta_{100}+\cos^2\xi\beta_{001})-1\right]. \label{eq:chi}
\end{align}
By reading off the coefficients of $\tilde{z}^2$ in Eq.~\eqref{eq:eulerstat}, we obtain a self-consistency relation for $R_z$:
\begin{equation}
  R_z^2 = \frac{2gn_0\zeta_z}{m\tilde{\omega}_z^2\omega_{\perp}^2} \equiv \left(\frac{15Na_{\text{s}}\zeta_z}{\kappa_x\kappa_y l_{\perp}\tilde{\omega}_z^2}\right)^{\frac{2}{5}}l_{\perp}^2. \label{eq:rzeqn}
\end{equation}
From the spatially constant terms in Eq.~\eqref{eq:eulerstat} we also find the chemical potential in the TF limit:
\begin{equation}
\mu = gn_0\chi \equiv \left(\frac{15Na_{\text{s}}}{\kappa_x\kappa_y l_{\perp}}\right)^{\frac{2}{5}}\left(\frac{\tilde{\omega}_z^2}{\zeta_z}\right)^{\frac{3}{5}}\frac{\hbar\chi\omega_{\perp}}{2}. \label{eq:chempotanswer}
\end{equation}
Similarly, the coefficients of $\tilde{x}^2$ and $\tilde{y}^2$ imply
\begin{equation}
    \kappa_i^2 = \frac{\tilde{\omega}_z^2\zeta_i}{\tilde{\omega}_i^2\zeta_z}. \label{eq:kappaeqn}
\end{equation}
By recognizing that there is no $\tilde{x}\tilde{z}$ term in Eq.~\eqref{eq:nstat}, we also obtain the condition that
\begin{equation}
S_{xz} = \frac{3\tilde{\omega}_z^2\varepsilon_{\text{dd}}\kappa_x\kappa_y\beta_{101}\sin(2\xi)}{\zeta_z} \label{eq:xzcoeffzero}
\end{equation}
By substituting Eq.~\eqref{eq:kappaeqn} into Eqs.~\eqref{eq:alphadefn} and \eqref{eq:deltadefn}, we obtain two final self-consistency relations, given by:
\begin{align}
    \alpha &= \left(\frac{\tilde{\omega}_y^2\zeta_x - \tilde{\omega}_x^2\zeta_y}{\tilde{\omega}_y^2\zeta_x + \tilde{\omega}_x^2\zeta_y}\right)\Omega\cos(\theta-\xi), \label{eq:alphaeqn} \\
    \delta &= \left(\frac{\tilde{\omega}_y^2\zeta_z - \tilde{\omega}_z^2\zeta_y}{\tilde{\omega}_y^2\zeta_z + \tilde{\omega}_z^2\zeta_y}\right)\Omega\sin(\theta-\xi), \label{eq:deltaeqn}
\end{align}

Since the solutions of Eqs.~\eqref{eq:kappaeqn} -- \eqref{eq:deltaeqn} can only have a physical meaning when the solutions for $R_i^2$ are real and positive, these self-consistency relations describe branches of stationary solutions that terminate at the points in parameter space when one or more of $\tilde{\omega}_x, \tilde{\omega}_y, \tilde{\omega}_z$ equal zero. These endpoints, and their values as functions of $\gamma$, $\varepsilon_{\text{dd}}$ and $\theta$, determine the number of stationary solutions that exist for a given value of $\Omega$. We find that the following limits are of use in attempting a self-consistent solution~\cite{pra_101_6_063638_2020}:
\begin{enumerate}[label=(\alph*)]
  \item $\tilde{\omega}_x \rightarrow 0$ and $\tilde{\omega}_y,\text{}\tilde{\omega}_z \neq 0$, \label{en:endpointx}
  \item $\tilde{\omega}_y \rightarrow 0$ and $\tilde{\omega}_x,\text{}\tilde{\omega}_z \neq 0$, \label{en:endpointy}
  \item $\tilde{\omega}_x,\text{}\tilde{\omega}_y \rightarrow 0$ and $\tilde{\omega}_z \neq 0$, \label{en:endpointxy}
  \item $\tilde{\omega}_y,\text{}\tilde{\omega}_z \rightarrow 0$ and $\tilde{\omega}_x \neq 0$. \label{en:endpointyz}
\end{enumerate}
For the sake of notational convenience we use the subscripts $xc$, $yc$, $xyc$, and $yzc$ to denote the values of various quantities in the limits~\ref{en:endpointx}, \ref{en:endpointy}, \ref{en:endpointxy} and \ref{en:endpointyz}, respectively. A detailed description of the self-consistency relations satisfied by $\Omega$, $\alpha$, $\delta$ and $\xi$ at each of these limits is provided in Appendix~\ref{sec:level6}, while a description of how the shape of the TF distribution can be understood via inspection of the signs of $\alpha$, $\delta$ and $\theta - \xi$ can be found in the literature~\cite{pra_101_6_063638_2020}. We also note that Eqs~~\eqref{eq:kappaeqn} -- \eqref{eq:deltaeqn} reduce to the special cases that have previously been analyzed in the literature in the appropriate limits. For instance, we recover the respective consistency relations for dipoles rotating about the rotation axis or in the plane with a normal given by the rotation axis by setting $\theta = \xi = 0$ or $\theta = \xi = \pi/2$, respectively~\cite{laserphys_18_5_322_2008, pra_80_3_033617_2009, prl_122_5_050401_2019, pra_100_2_023625_2019}. In both instances, Eq.~\eqref{eq:deltaeqn} has the trivial solution $\delta = 0$ and both sides of Eq.~\eqref{eq:xzcoeffzero} are zero. Similarly, the self-consistency relations obtained by setting $\Omega = 0$, which results in Eqs.~\eqref{eq:alphaeqn} and \eqref{eq:deltaeqn} yielding the trivial solutions $\alpha = \delta = 0$, are identical to those in the literature for non-rotating dipoles aligned at an arbitrary angle to the symmetry axes of an anisotropic, nonrotating harmonic trap~\cite{pra_82_5_053620_2010}.

\section{\label{sec:level3}Stationary Solution Branches}
\subsection{\label{sec:level3a}Slow and Intermediate Rotation Frequencies}
\begin{figure*}[ht]
\includegraphics[width=\linewidth]{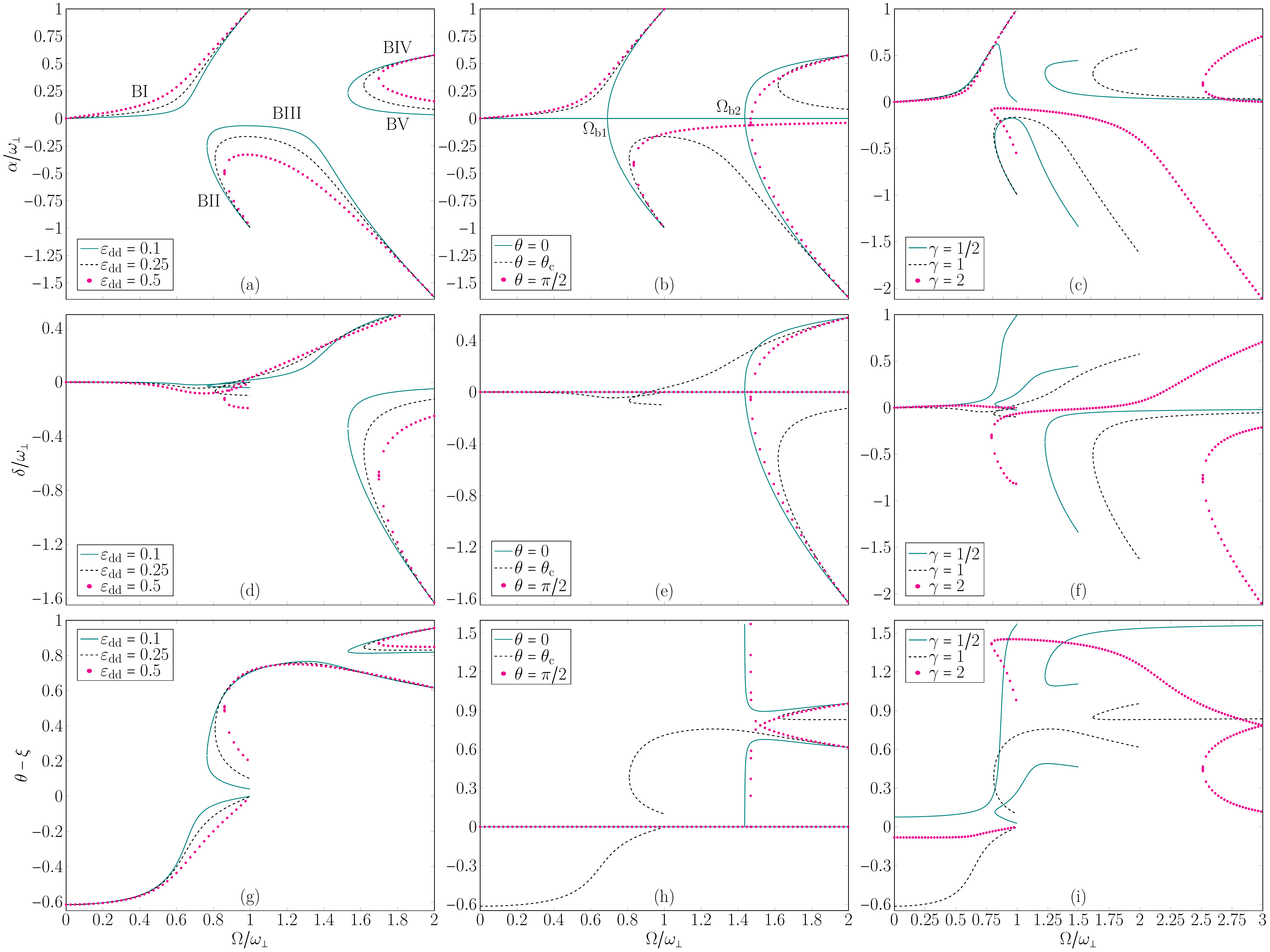}
\vspace*{-5mm}
\caption{Stationary solutions in a rotating reference frame for $\alpha$ (first row, in units of $\omega_{\perp}$), $\delta$ (second row, in units of $\omega_{\perp}$), and $\theta - \xi$ (third row, in radians), as functions of $\Omega$ (in units of $\omega_{\perp}$), demonstrating the existence of two distinct bifurcations and a dependence of the branch endpoints on $\gamma$, rather than $\varepsilon_{\text{dd}}$ or $\theta$. The first column fixes $\gamma = 1,\,\theta = \theta_{\text{c}}$, whereas the second column fixes $\gamma = 1,\,\varepsilon_{\text{dd}} = 0.25$, and the third column fixes $\varepsilon_{\text{dd}} = 0.25,\,\theta = \theta_{\text{c}}$. The branches I -- V are denoted as BI -- BV in (a), while the bifurcation frequencies $\Omega_{\text{b}1}$ and $\Omega_{\text{b}2}$ are labeled in (b) for reference.}
\label{slowrotations}
\end{figure*}

We now proceed to present the semi-analytical solutions of Eqs.~\eqref{eq:continuity} and \eqref{eq:euler} as defined through Eqs.~\eqref{eq:kappaeqn} -- \eqref{eq:deltaeqn}. In Appendix~\ref{sec:level6} we show that $\Omega_{xyc} = \Omega_{yzc} = (1 + \gamma)\omega_{\perp}$ and that $\Omega_{xc} = \Omega_{yc} = \omega_{\perp}$. Therefore, the stationary solutions for the rotation of the dipole moments may be classified into two related regimes based on the value of $\Omega$; in this section we present a discussion of the stationary solutions of Eqs.~\eqref{eq:continuity} and \eqref{eq:euler} for $0 \leq \Omega \leq (1 + \gamma)\omega_{\perp}$, and in Section~\ref{sec:level3b} we analyze the stationary solutions in the regime where $\Omega \gg (1 + \gamma)\omega_{\perp}$. To provide a broadly representative sample of the possible stationary solution regimes for a given value of $\Omega$, we restrict our analysis to the following cases:
\begin{enumerate}
\item $\gamma = 1$, $\varepsilon_{\text{dd}} = \lbrace 1/10, 1/4, 1/2\rbrace$, $\theta = \theta_{\text{c}}$.
\item $\gamma = 1$, $\varepsilon_{\text{dd}} = 1/4$, $\theta \in \lbrace 0, \theta_{\text{c}}, \pi/2\rbrace$.
\item $\gamma = \lbrace 1/2, 1, 2\rbrace$, $\varepsilon_{\text{dd}} = 1/4$, $\theta = \theta_{\text{c}}$.
\end{enumerate}
In Fig.~\ref{slowrotations} we present plots of $\alpha$ (1st row, in units of $\omega_{\perp}$), $\delta$ (2nd row, in units of $\omega_{\perp}$), and $\theta - \xi$ (3rd row) as functions of $\Omega$ (in units of $\omega_{\perp}$) for these three cases, with the $i$th case presented in the $i$th column.

When $\Omega = 0$, the stationary states have already been described in the literature: the semi-axes along the dipole orientation are larger than their respective values for a nondipolar BEC in an identical trap, an effect referred to as magnetostriction~\cite{pra_71_3_033618_2005, pra_82_3_033612_2010}. Furthermore, when $\gamma \neq 1$ and $\sin(2\theta) \neq 0$, the value of $\xi$ is a nontrivial function of $\varepsilon_{\text{dd}}$, $\gamma$, and $\theta$, reflecting the competitive interplay between the external trapping potential and the internal dipole-dipole interaction on the density profile of the condensate~\cite{pra_82_5_053620_2010}.

The salient features of the stationary solution structure can be observed in Figs.~\ref{slowrotations} (a) -- (c), which plot $\alpha$ as a function of $\Omega$. When one or more of $\varepsilon_{\text{dd}}$ and $\sin(2\theta)$ are zero, there always exists a solution such that $\alpha = 0$, a consequence of the rotational symmetry of the system about the $z$-axis. This is also reflected in the corresponding solutions, $\delta = \theta - \xi = 0$, in the second and third rows of Fig.~\ref{slowrotations} since the absence of a tilted potential allows the condensate density to exhibit zero tilting about the rotation axis. If $\varepsilon_{\text{dd}}$ and $\sin(2\theta)$ are both nonzero, we find that $\alpha$ and $\delta$ vanish only when $\Omega = 0$. Instead of a constant $\alpha = 0$ solution, a stationary solution branch featuring a nonzero value of $\alpha$ exists for the interval $0 \leq \Omega < \omega_{\perp}$. In the limit $\Omega \rightarrow \Omega_{xc} = \omega_{\perp}$ we find that this branch, which we refer to as branch I and is denoted as BI in Fig.~\ref{slowrotations} (a), satisfies $\alpha \rightarrow \alpha_{xc} = \omega_{\perp}$, $\delta \rightarrow \delta_{xc} = 0$, and $\xi \rightarrow \xi_{xc} = \theta$. As elucidated in Appendix~\ref{sec:level6}, the body frame axis may be rotated about the $\tilde{y}$-axis by $\pi/2$ without loss of generality and the branch endpoint is subsequently characterized by $\alpha \rightarrow 0$, $\delta_{xc} \rightarrow \omega_{\perp}$, and $\xi \rightarrow \theta - \pi/2$, when $\Omega \rightarrow \omega_{\perp}$. This behavior is consistent with that of the stationary states that have been described in the literature for the polarization tilting angles $\theta = 0$ and $\theta = \pi/2$~\cite{laserphys_18_5_322_2008, pra_80_3_033617_2009, prl_122_5_050401_2019, pra_100_2_023625_2019}. Another characteristic feature of the stationary solutions that is in agreement with these previous studies is that a second stationary solution branch, which we refer to as branch II and denote in Fig.~\ref{slowrotations} (a) as BII, begins at a bifurcation frequency $\Omega = \Omega_{\text{b}1}$ and terminates in the limit $\Omega = \Omega_{yc} \equiv \omega_{\perp}$. Note that $\Omega_{\text{b}1}$ is labeled in Fig.~\ref{slowrotations} (b). When $\sin(2\theta) = 0$, $\alpha \rightarrow \alpha_{yc} = -1$, $\delta \rightarrow \delta_{yc} = 0$, and $\xi \rightarrow \xi_{yc} = \theta$, but for nonzero $\sin(2\theta)$ these limits must be determined from a set of self-consistency relations that are presented in Appendix~\ref{sec:level6}.

Another feature that has previously been noted in the $\sin(2\theta) \rightarrow 0$ limit is the existence of a third stationary solution branch, branch III, that begins at $\Omega = \Omega_{\text{b}1}$; it is denoted as BIII in Fig.~\ref{slowrotations} (a). If $\sin(2\theta) = 0$, this branch persists for all $\Omega \geq \Omega_{\text{b}1}$ while both $\alpha$ and $\delta$ approach zero as $\Omega \rightarrow \infty$~\cite{pra_80_3_033617_2009, pra_100_2_023625_2019}. However, the behavior of branch III is very different when $\sin(2\theta) \neq 0$ and instead resembles that of the analogous branch for a nondipolar condensate in a tilted, rotating harmonic trap~\cite{pra_101_6_063638_2020}. More specifically, the branch does not extend till infinity but rather terminates when $\Omega \rightarrow \Omega_{xyc} \equiv (1 + \gamma)\omega_{\perp}$, with the respective values of $\tilde{\alpha}$, $\tilde{\delta}$, and $\theta - \xi$ in this limit being specified in Appendix~\ref{sec:level6}. This unusual behavior is accompanied by the existence of two additional stationary state branches, connected to each other at the bifurcation frequency $\Omega = \Omega_{\text{b}2}$, regardless of the choice of $\gamma$, $\theta$ or $\varepsilon_{\text{dd}}$. Note that, like $\Omega_{\text{b}1}$, we have labeled $\Omega_{\text{b}2}$ in Fig.~\ref{slowrotations} (b).

If $\sin(2\theta) = 0$, the branches emerging at $\Omega = \Omega_{\text{b}2}$ terminate when $\Omega = \Omega_{xyc} = \Omega_{yzc} = (1 + \gamma)\omega_{\perp}$, as seen in the second column of Fig.~\ref{slowrotations}. They are also characterized by a nonzero body frame axis tilt, $\theta - \xi$, despite the null tilting of the dipole polarization. Indeed, we find in Fig.~\ref{slowrotations} (h) that the sum of $\theta - \xi$ for these two branches is $\pi/2$, and in Fig.~\ref{slowrotations} (b) and (h) that the value of $\alpha$ for one of the branches is equal to that of $\delta$ for the other branch, and vice versa. Equations~\eqref{eq:alphadefn} and \eqref{eq:deltadefn} therefore imply that the two branches exhibit identical values of $R_y$, with the value of $R_x$ for one of the branches equaling that of $R_z$ for the other. Since the sum of the two values of $\theta - \xi$ is $\pi/2$, the two stationary solutions correspond to identical TF densities that are rotated at equal, but opposite, angles about the $\tilde{y}$-axis. However, this equivalence does not exist for nonzero $\sin(2\theta)$. Instead, one of the branches, which we refer to as branch IV, terminates at $\Omega = \Omega_{yzc} = (1 + \gamma)\omega_{\perp}$ as seen in the third column of Fig.~\ref{slowrotations}. Conversely, the fifth branch, which we refer to as branch V, persists for all $\Omega \geq \Omega_{\text{b}2}$ and is characterized by $\alpha$ and $\delta$ vanishing in the limit $\Omega \rightarrow \infty$. Note that branches IV and V are denoted in Fig.~\ref{slowrotations} as BIV and BV, respectively.

In Fig.~\ref{slowrotations} it is evident that, regardless of the values of $\theta$, $\gamma$, or $\varepsilon_{\text{dd}}$, there exist five distinct stationary solution branches and two distinct bifurcations. Let us consider the nature of the bifurcations more closely. Their existence in Fig.~\ref{slowrotations} may be attributed to the energetic instability of quadrupolar surface modes, induced by the rotation of the reference frame, that permit elliptical deformations of the condensate density. Let us denote the inertial-frame value of the angular frequency of a surface mode with azimuthal and magnetic quantum numbers $l$ and $m_z$, respectively, as $\omega_{\text{s}}(l, m_z)$. The value of the angular frequency of this mode, as measured in a frame rotating about the magnetic axis with an angular frequency $\Omega$ is then given by $\omega_{\text{r}}(l, m_z) = \omega_{\text{s}}(l, m_z) - |m_z|\Omega$ and so the condensate is unstable against this mode when $\omega_{\text{r}}(l, m_z) < 0$, i.e. $ \Omega > \omega_{\text{s}}(l, m_z)/|m_z|$~\cite{pra_63_1_011601r_2000}. In previous studies of the $\theta = 0$ limit the first bifurcation, which introduces an ellipticity of the density profile about the $z$-axis, has been found to be the result of the instability of the $l = 2,\,m_z = \pm 2$ surface modes. Therefore when $\theta = 0$, the first bifurcation frequency $\Omega_{\text{b}1}$ satisfies the relation $\omega_{\text{s}}(2, \pm 2) - 2\Omega_{\text{b}1} = 0$. In the nondipolar limit, $\omega_{\text{s}}(2, \pm 2) = \sqrt{2}\omega_{\perp}$ and thus $\Omega_{\text{b}1} = \omega_{\perp}/\sqrt{2}$ when $\varepsilon_{\text{dd}} = 0$~\cite{prl_77_12_2360-2363_1996, prl_86_3_377-380_2001}; for nonzero $\varepsilon_{\text{dd}}$ the value of $\omega_{\text{s}}(2, \pm 2)$ does not have a closed form and must be evaluated via the numerical solution of self-consistency relations for the mode frequency~\cite{pra_82_3_033612_2010, pra_82_5_053620_2010}. Similarly, the second bifurcation at $\Omega = \Omega_{\text{b}2}$, which is intrinsically associated with a tilting of the condensate density profile about the $\tilde{y}$-axis, is associated with the scissors modes $l = 2,\,m_z = \pm 1$. This rotating-frame instability has recently been found to manifest itself in nondipolar condensates in rotating harmonic traps~\cite{pra_101_6_063638_2020}, and the presence of the associated bifurcation in a dipolar condensate with rotating dipole moments is not surprising. We also note that similar bifurcations are seen in other systems such as an irrotational gravitationally-bound fluid that can transform from a Maclaurin spheroid to a tilted Riemann ellipsoid~\cite{physfluids_8_12_3414-3422_1996}. The instability of the scissors mode results in a bifurcation frequency $\Omega_{\text{b}2} = \sqrt{1 + \gamma^2}\omega_{\perp}$ when $\varepsilon_{\text{dd}} = 0$~\cite{prl_77_12_2360-2363_1996}. For nonzero $\varepsilon_{\text{dd}}$ with $\theta = 0$, the self-consistency relations that define the mode frequency are found elsewhere~\cite{pra_82_3_033612_2010, pra_82_5_053620_2010}. However, we note that the method of determination of the bifurcation frequencies through examination of the condensate's surface mode spectrum is only valid in the presence of axial symmetry in the density profile about the rotation axis, given that $m_z$ is no longer a good quantum number in its absence.

By allowing for the condensate density's body frame to be tilted with respect to the trap's symmetry axes, we have uncovered two new stationary state branches that exist regardless of the tilting angle of the dipole moments. We now address the question of how any one of the five branches may be accessed in either experiments or numerical simulations. In a quasi-adiabatic rampup of the rotation frequency from zero at a nonzero value of $\varepsilon_{\text{dd}}$, the condensate would be expected to follow branch I~\cite{laserphys_18_5_322_2008, pra_80_3_033617_2009, pra_100_2_023625_2019}. However, from an inspection of the second column of Fig.~\ref{slowrotations}, it is possible to access branch III by starting in a rotating reference frame, with the dipole moments aligned along the rotation axis, and quasi-adiabatically increasing the value of $\theta$ from zero~\cite{prl_122_5_050401_2019}. At a final value of $\theta$, halting the quasi-adiabatic tuning of $\theta$ and initiating a similar tuning procedure for $\Omega$ would allow experimentalists to access this branch at different rotation frequencies. Throughout, it would be necessary to ensure that the rotation frequency is higher than the first bifurcation frequency at the current value of $\theta$; if the current value of $\Omega_{\text{b}1}$ falls below the rotation frequency, the absence of a connected stationary solution branch would destabilize the condensate. Similarly, if the final value of $\theta$ is less than $\pi/2$, this branch does not extend to infinity but terminates at the rotation frequency $\Omega = (1 + \gamma)\omega_{\perp}$ and any rotation frequency tuning procedure would need to avoid this terminal. Instead, if the quasi-adiabatic rampup of $\theta$ from zero is initialized at values of $\Omega$ higher than $\Omega_{\text{b}2}$, the stationary solution branch that are accessed by the condensate is always branch V~\cite{pra_101_4_043606_2020}. As such, we predict that at least three out of the five stationary solution branches may readily be accessed through parameter tuning protocols; it may not be possible to access branches II and IV via such procedures.

Finally, we note the effects of altering one of $\varepsilon_{\text{dd}}$, $\theta$ or $\gamma$ with the remaining pair of parameters being fixed. In the first column of Fig.~\ref{slowrotations} we find that an increase in the value of $\varepsilon_{\text{dd}}$ results in higher values of both $\Omega_{\text{b}1}$ and $\Omega_{\text{b}2}$ but does not alter the stationary solutions qualitatively. Conversely, three of the five stationary solution branches are found to be untilted ($\theta - \xi = 0$) when $\sin(2\theta) = 0$, while all five branches feature nonzero values of $\theta - \xi$ for any other value of $\theta$, as seen in the second column of Fig.~\ref{slowrotations}. In the third column of the figure we see that changing the value of $\gamma$ changes the values of $\Omega_{xyc}$ and $\Omega_{yzc}$, two of the four branch endpoints, which are equivalent to $(1 + \gamma)\omega_{\perp}$. The qualitative properties of the five branches are otherwise not affected by changes in the value of $\gamma$. While branches I and II in Figs.~\ref{slowrotations} (c) and (f) appear to be qualitatively distinct for $\gamma = 1/2$ as compared to $\gamma = \lbrace 1,\,2\rbrace$, this is due to a rotation in $\tilde{\mathbf{r}}$ by $\pi/2$ about the $\tilde{y}$-axis that we have made so that the branches for $\gamma = 1/2$ do not overlap with each other.

\subsection{\label{sec:level3b}Ultrarapid Rotation Frequencies}
\begin{figure*}[ht]
\includegraphics[width=\linewidth]{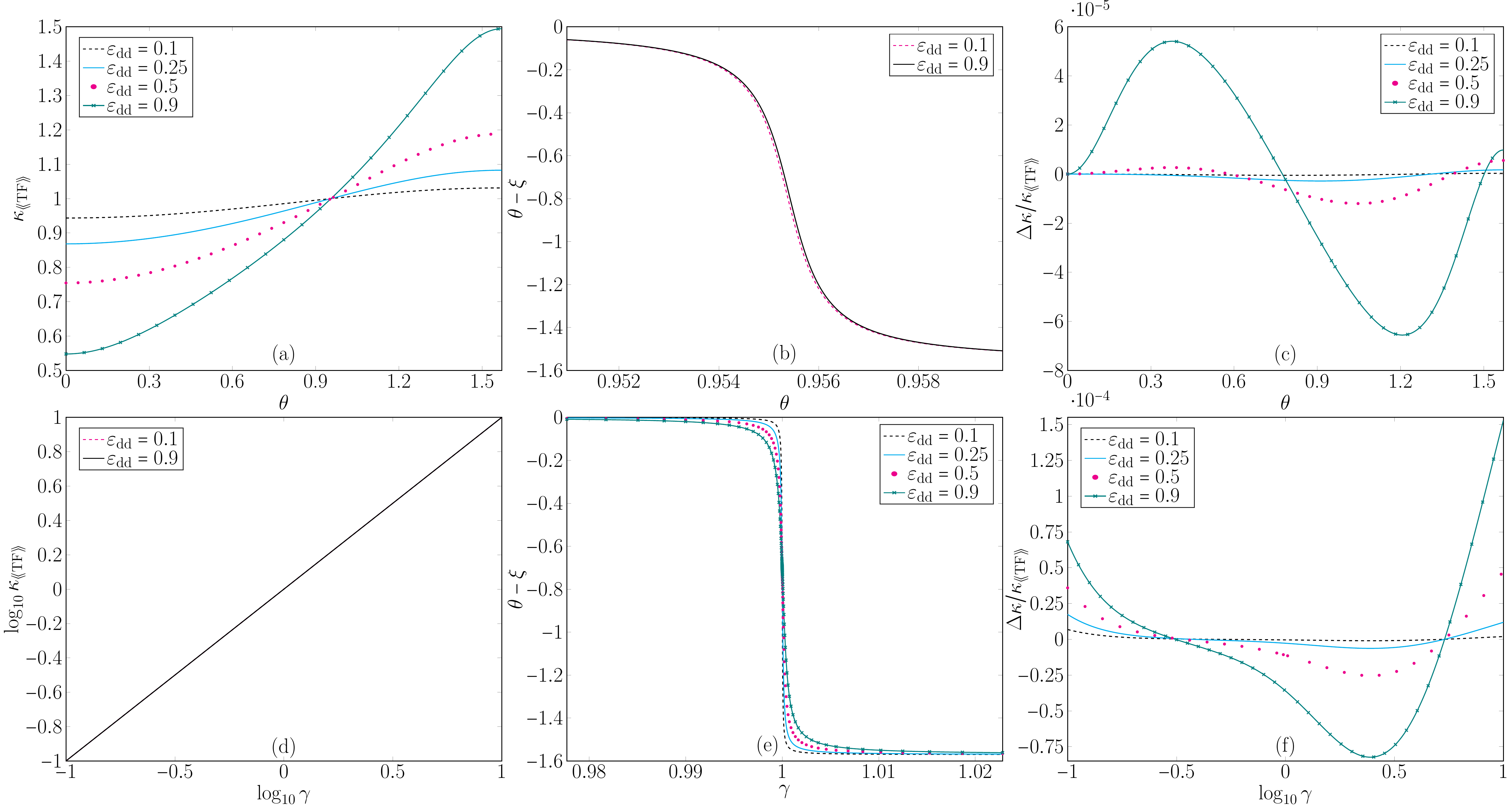}
\vspace*{-5mm}
\caption{Stationary solutions for $\Omega = 50\omega_{\perp}$: in the first row, $\theta\in[0, \pi/2]$ with fixed $\gamma = 1$ whereas in the second row, $\gamma\in[1/10,10]$ with fixed $\theta = \theta_{\text{c}} = \arccos(1/\sqrt{3}) \approx 0.955$. The first column represents $\kappa_{\llangle\text{TF}\rrangle}$ as defined in the main text, the second column represents $\theta - \xi$ in radians, and the third column represents the relative difference between the time-averaged aspect ratio and the aspect ratio from the time-averaged DDI.}
\label{fastrotations}
\end{figure*}

For the regime where $\Omega\in[0, 1 + \gamma)\omega_{\perp}$, we have seen that there exist five distinct branches of stationary solutions in the TF limit. For rotation frequencies greater than $(1 + \gamma)\omega_{\perp}$, Figs.~\ref{slowrotations} (c), (f) and (i) predict that the solution of Eqs.~\eqref{eq:contstat} and \eqref{eq:eulerstat} is unique. These stationary solutions are characterized by an ever-increasing degree of axial symmetry of the condensate density about the rotation axis for larger rotation frequencies since $\alpha$ and $\delta$ tend to zero as $\Omega\rightarrow\infty$. Due to the axial symmetry of $V_{\text{T}}$ and $\llangle U_{\text{dd}}(\mathbf{r})\rrangle$ about the rotation axis, this observation is consistent with the stationary solutions in this limit being governed by the effective dipolar interaction, $\llangle U_{\text{dd}}(\mathbf{r})\rrangle$. Given that Eq.~\eqref{eq:DDIkerneltimeavg} is predicted to be valid for $\Omega \gg \omega_{\perp}$, it is prudent to verify the consistency of the stationary solutions with the time-averaging scheme for dipole rotation frequencies of at least an order of magnitude larger than $\omega_{\perp}$. Therefore we proceed by solving Eqs.~\eqref{eq:kappaeqn} -- \eqref{eq:deltaeqn} for a range of different values of $\gamma$, $\varepsilon_{\text{dd}}$, and $\theta$, with the dipole rotation frequency $\Omega = 50\omega_{\perp}$. In Fig.~\ref{fastrotations} (a) -- (c) we have presented the results of varying $\theta$ from $0$ to $\pi/2$ in a spherical trap ($\gamma = 1$) for multiple values of $\varepsilon_{\text{dd}}$, given that in recent numerical and experimental investigations of the rotational tuning scheme, the dipole polarization angle was varied in this manner~\cite{prl_120_23_230401_2018, pra_101_4_043606_2020}. We have also investigated the effects fixing $\theta = \theta_{\text{c}}$, such that Eq.~\eqref{eq:DDIkerneltimeavg} vanishes, and varying the trapping aspect ratio from strongly prolate ($\gamma = 1/10$) to oblate ($\gamma = 10$) for multiple values of $\varepsilon_{\text{dd}}$. The results of this procedure are given in Fig.~\ref{fastrotations} (d) -- (f).

Figures~\ref{fastrotations} (a) and (d) plot $\kappa_{\llangle\text{TF}\rrangle}$, the time-averaged ratio of the in-plane TF semi-axis to the TF semi-axis along the rotation axis. To derive this quantity, we consider the TF density profile corresponding to a stationary solution in the body frame, $n_{\text{TF}}(\tilde{\mathbf{r}})$, and define the upright laboratory frame coordinates, $\mathbf{r}'$, such that
\begin{equation}
    \begin{pmatrix}
    x \\
    y \\
    z \\
    \end{pmatrix}
    =
    \begin{pmatrix}
    \cos(\Omega t) & \sin(\Omega t) & 0 \\
    -\sin(\Omega t) & \cos(\Omega t) & 0 \\
    0 & 0 & 1
    \end{pmatrix}
    \begin{pmatrix}
    x' \\
    y' \\
    z' \\
    \end{pmatrix}. \label{eq:labtocorot}
\end{equation}
Via a short calculation, we find via Eqs.~\eqref{eq:dipoleadjusttiltframe}, \eqref{eq:nstat} and \eqref{eq:labtocorot} that the time-average of the TF density profile over one rotation cycle of the dipole polarization is given by
\begin{align}
    \llangle n(\mathbf{r}')\rrangle &= n_0\left(1 - \frac{\rho'^2}{\llangle R_{\rho}\rrangle^2} - \frac{z'^2}{\llangle R_z\rrangle^2}\right), \label{eq:nstattimeavg} \\
    \llangle R_{\rho}\rrangle^2 &= \frac{2\kappa_x^2\kappa_y^2R_z^2}{\kappa_x^2 + \kappa_y^2\cos^2(\theta-\xi) + \kappa_x^2\kappa_y^2\sin^2(\theta-\xi)}, \label{eq:rrho2timeavg} \\
    \llangle R_{z}\rrangle^2 &= \frac{\kappa_x^2R_z^2}{\sin^2(\theta-\xi) + \kappa_x^2\cos^2(\theta-\xi)}. \label{eq:rz2timeavg}
\end{align}
The time-averaged aspect ratio of the condensate density in the plane with respect to the rotation axis, $\kappa_{\llangle\text{TF}\rrangle}$, is therefore specified by
\begin{equation}
    \kappa_{\llangle\text{TF}\rrangle}^2 = \frac{2\kappa_y^2\left[\sin^2(\theta-\xi) + \kappa_x^2\cos^2(\theta-\xi)\right]}{\kappa_x^2 + \kappa_y^2\cos^2(\theta-\xi) + \kappa_x^2\kappa_y^2\sin^2(\theta-\xi)}. \label{eq:kappaperptfavg}
\end{equation}

We also define $\kappa_{\llangle\text{DDI}\rrangle}$ as the corresponding aspect ratio satisfied by the stationary state for non-rotating dipole moments that experience the time-averaged DDI, Eq.~\eqref{eq:DDIkerneltimeavg}. Equations \eqref{eq:omegaxeff} -- \eqref{eq:deltadefn} imply that $\kappa_{\llangle\text{DDI}\rrangle}$ obeys the self-consistency relation,
\begin{equation}
\kappa_{\llangle\text{DDI}\rrangle}^2 = \gamma^2\left[\frac{4 + \varepsilon_{\text{dd}}(3\cos^2\theta - 1)\left(3\kappa_{\llangle\text{DDI}\rrangle}^4\beta_{101} - 2\right)}{4 + \varepsilon_{\text{dd}}(3\cos^2\theta - 1)\left(9\kappa_{\llangle\text{DDI}\rrangle}^2\beta_{002} - 2\right)}\right], \label{eq:kapparottunedoriginal}
\end{equation}
where both of the arguments of $\beta_{101}$ and $\beta_{200}$ are equal to $\kappa_{\llangle\text{DDI}\rrangle}$. By evaluating these integrals, and defining $\overline{\varepsilon}_{\text{dd}} = \varepsilon_{\text{dd}}(3\cos^2\theta-1)/2$, one may show that Eq.~\eqref{eq:kapparottunedoriginal} is equivalent to the relation,
\begin{align}
    \kappa_{\llangle\text{DDI}\rrangle}^2 &= \frac{2\gamma^2(1-\overline{\varepsilon}_{\text{dd}})}{2+\overline{\varepsilon}_{\text{dd}}\left[4-3(\gamma^2+2)F(\kappa_{\llangle\text{DDI}\rrangle})\right]}, \label{eq:kapparottuned} \\
    F(\kappa) &= \frac{1 + 2\kappa^2}{(1 - \kappa^2)^2} - \frac{3\kappa^2\arctanh({\sqrt{1 - \kappa^2}})}{(1 - \kappa^2)^{5/2}}, \label{eq:dipolarfunction}
\end{align}
which has been derived in previous studies of the non-rotating, axially symmetric dipolar TF ground state~\cite{prl_92_25_250401_2004, pra_71_3_033618_2005, pra_74_1_013621_2006, pra_86_6_063609_2012}. Therefore, for rotation frequencies of orders of magnitude larger than the trapping frequency, such as in Fig.~\ref{fastrotations}, we expect that $\kappa_{\llangle\text{TF}\rrangle} \approx \kappa_{\llangle\text{DDI}\rrangle}$.

Figure~\ref{fastrotations} demonstrates that this is indeed the case. In a spherically symmetric trap, the time-averaged condensate density experiences magnetostriction along the rotation axis for $\theta < \theta_{\text{c}}$ and the surfaces of constant density are thus prolate spheroids, as seen in Fig.~\ref{fastrotations} (a) where $\kappa_{\llangle\text{TF}\rrangle} < 1$. We also see that for larger values of $\varepsilon_{\text{dd}}$, the degree of magnetostriction is correspondingly larger. When the dipole orientation angle is equal to $\theta_{\text{c}} = \arccos{1/\sqrt{3}} \approx 0.955$, the time-averaged condensate density is spherically symmetrical for any value of $\varepsilon_{\text{dd}}$. As the dipole tilt, $\theta$, increases further towards $\pi/2$, the time-averaged condensate density experiences an inverse magnetostrictive effect where the surfaces of constant density are oblate spheroids, such that $\kappa_{\llangle\text{TF}\rrangle} > 1$. We also find that $\kappa_{\llangle\text{TF}\rrangle}$ is larger for increasing $\varepsilon_{\text{dd}}$. These features agree with the predicted behavior of $\kappa_{\llangle\text{DDI}\rrangle}$~\cite{prl_89_13_130401_2002, pra_71_3_033618_2005, pra_82_3_033612_2010} and, by explicitly solving Eqs.~\eqref{eq:contstat} and \eqref{eq:eulerstat} in the co-rotating frame for a finite rotation frequency and time-averaging the resulting density profile, we have uncovered the role played by the interplay between the dipolar interaction and centrifugal distortion in ensuring that the time-averaging property is satisfied. We also plot the relative difference between $\kappa_{\llangle\text{TF}\rrangle}$ and $\kappa_{\llangle\text{DDI}\rrangle}$,
\begin{equation}
    \frac{\Delta\kappa}{\kappa_{\llangle\text{TF}\rrangle}} \equiv \frac{\kappa_{\llangle\text{TF}\rrangle} - \kappa_{\llangle\text{DDI}\rrangle}}{\kappa_{\llangle\text{TF}\rrangle}}, \label{eq:reldiff}
\end{equation}
in Fig.~\ref{fastrotations} (c) and find that it is of the order of $10^{-5}$ at the most. Similarly, the prediction of Eq.~\eqref{eq:kapparottuned} that $\kappa_{\llangle\text{TF}\rrangle} \approx \gamma$ when $\theta = \theta_{\text{c}}$ is supported by Fig.~\ref{fastrotations} (d), where the dipole tilt is fixed at $\theta = \theta_{\text{c}}$ and the trap aspect ratio $\gamma$ is varied. Furthermore, the individual lines for $\varepsilon_{\text{dd}} = 0.1$ (magenta, dashed) and $\varepsilon_{\text{dd}} = 0.9$ (black, solid) are almost entirely incident upon each other, a consequence of the universality of the `magic angle', $\theta_{\text{c}}$, with respect to $\varepsilon_{\text{dd}}$. This is supported by Fig.~\ref{fastrotations} (f) in which the corresponding plots of $\Delta\kappa/\kappa_{\llangle\text{TF}\rrangle}$ versus $\log_{10}\gamma$, for distinct values of $\varepsilon_{\text{dd}}$, suggest that the relative difference between $\kappa_{\llangle\text{TF}\rrangle}$ and $\kappa_{\llangle\text{DDI}\rrangle}$ is at most of the order of $10^{-4}$.

Figure~\ref{fastrotations} also reveals that for dipole tilting angles and/or trapping aspect ratios far from the thresholds, $\theta \simeq \theta_{\text{c}}$ and $\gamma \simeq 1$, the condensate density is almost perfectly aligned along the untilted rotating frame axes, i.e. $\sin(\theta - \xi) \approx 0$. When sweeping over $\theta$ from $0$ to $\pi/2$, or when sweeping over $\gamma$ from $\gamma \ll 1$ to $\gamma \gg 1$, we observe a rapid transition of the density's tilt angle from $\theta - \xi \approx 0$ to $\theta - \xi \approx -\pi/2$; these transitions are in the vicinity of $\theta \simeq \theta_{\text{c}}$ and $\gamma \simeq 1$, respectively. In Fig.~\ref{fastrotations} (b) we see that when sweeping over $\theta$ at constant $\gamma = 1$, the transition from $\theta - \xi \approx 0$ to $\theta - \xi \approx -\pi/2$ is not sensitive to the value of $\varepsilon_{\text{dd}}$ and that the respective curves for $\varepsilon_{\text{dd}} = 0.1$ and $\varepsilon_{\text{dd}} = 0.9$ are almost identical. This is in contrast to Fig.~\ref{fastrotations} (e), which plots $\theta - \xi$ as a function of $\gamma$ near $\gamma = 1$ for constant $\theta = \theta_{\text{c}}$, where it is clearly evident that the transition from $\theta - \xi = 0$ to $\theta - \xi = \pi/2$ is faster for smaller values of $\varepsilon_{\text{dd}}$. Combined with the property that the graphs in Fig.~\ref{fastrotations} (a) are clearly distinguishable from each other whereas those in Fig.~\ref{fastrotations} (d) are not, this would indicate that the condensate's properties are much more sensitive to the trapping geometry than the dipolar interaction in the ultrarapid rotation limit.

\section{\label{sec:level4}Dynamical Stability at Ultrarapid Rotation Frequencies}
Through the equations of superfluid quantum hydrodynamics that were set up in Sec.~\ref{sec:level1} and solved in \ref{sec:level2}, we have shown in Sec.~\ref{sec:level3a} that the continuous rotation of the dipole moments can induce nontrivial tilting angles of the condensate's TF stationary state density. In addition, the results of Sec.~\ref{sec:level3b} demonstrate that the time-averaged dipolar interaction effectively governs the properties of the TF stationary states when $\Omega$ is at least an order of magnitude greater than $\omega_{\perp}$. However, it has been established in previous semi-analytical and numerical studies that when $\theta = \pi/2$ and $\Omega > (1 + \gamma)\omega_{\perp}$, the corresponding stationary solution plotted in Fig.~\ref{slowrotations} (b) suffers from a dynamical instability that causes a dipolar condensate to undergo turbulent decay from the Thomas-Fermi state~\cite{pra_100_2_023625_2019}. Indeed, numerical simulations of the dGPE have predicted that the instability of this stationary solution can occur whenever the dipole alignment is not parallel to the rotation axis~\cite{prl_122_5_050401_2019, pra_101_4_043606_2020}. This would limit the efficacy of rapidly rotating the dipole polarization to effectively tune the DDI since the lifetime of the states described in Sec.~\ref{sec:level3b} would not be long enough for meaningful experimental analyses to be conducted. Therefore, as a preliminary investigation of the dynamical instability, we devote this section to obtaining the spectrum of collective excitations of a dipolar condensate in the TF regime for $\Omega > \Omega_{\text{b}2}$.

The collective excitations of a dipolar BEC are well-described in the linear regime by the time-dependent superfluid hydrodynamic equations, Eqs.~\eqref{eq:continuity} and \eqref{eq:euler}, linearized about their stationary solutions. In this formalism, which is valid when the time-dependent perturbations of the system are of sufficiently small magnitude, the collective modes are expressed as time-dependent fluctuations of the density and the phase, which are together equivalent to the solutions of the Bogoliubov--de Gennes equations~\cite{pra_54_5_4204-4212_1996, pra_58_4_3168-3179_1998, pitaevskiistringaribec}. In order to determine the collective mode spectrum, we may write
\begin{align}
n(\tilde{\mathbf{r}}, t) &= n_{\text{TF}}(\tilde{\mathbf{r}}) + \delta n(\tilde{\mathbf{r}}, t), \label{eq:densitypert} \\
S(\tilde{\mathbf{r}}, t) &= S_{\text{TF}}(\tilde{\mathbf{r}}, t) + \delta S(\tilde{\mathbf{r}}, t), \label{eq:phasepert}
\end{align}
where $S_{\text{TF}}(\tilde{\mathbf{r}}, t) = -\mu t/\hbar + \alpha\tilde{x}\tilde{y} + \delta\tilde{y}\tilde{z}$ and $\lbrace\delta n,\,\delta S\rbrace$ represent the collective modes. The subsequent linearization of Eqs.~\eqref{eq:continuity} and \eqref{eq:euler} is achieved through neglecting quadratic contributions from $\delta n$ and $\delta S$. This results in a coupled set of first-order equations for the time-evolution of the modes, given by~\cite{prl_87_19_190402_2001, pra_73_6_061603r_2006}
\begin{gather}
\frac{\partial}{\partial t}
\begin{pmatrix}
\delta S \\
\delta n
\end{pmatrix}
=
\mathcal{M}
\begin{pmatrix}
\delta S \\
\delta n
\end{pmatrix}, \label{eq:perteqns} \\
\mathcal{M} = -\begin{pmatrix}
\mathbf{v}_c\cdot\nabla & \frac{g}{\hbar}(1 - \varepsilon_{\text{dd}}\widehat{K}) \\
\frac{\hbar}{m}\nabla\cdot\left(n_{\text{TF}}\nabla\right) & \mathbf{v}_c\cdot\nabla
\end{pmatrix}, \label{eq:pertmatrix} \\
\mathbf{v}_c = \frac{\hbar}{m}\nabla S_{\text{TF}} - \mathbf{\Omega}\times\tilde{\mathbf{r}}, \label{eq:labvel} \\
\widehat{K}\left[f(\tilde{\mathbf{r}}, t)\right] = f(\tilde{\mathbf{r}}, t) + 3\left(\widehat{\mathbf{B}}\cdot\nabla\right)^2\int_{\Gamma_{\text{TF}}}\frac{f\left(\tilde{\mathbf{r}}',t\right)\,\mathrm{d}^3\tilde{r}'}{4\pi\left|\tilde{\mathbf{r}}-\tilde{\mathbf{r}}'\right|}, \label{eq:koperator}
\end{gather}
where $\Gamma_{\text{TF}}$, the domain of the integral in Eq.~\eqref{eq:koperator}, is defined as the region where $n_{\text{TF}}$ is nonzero. Hence, we write the density fluctuation of the $\nu$th collective mode as $\delta n_{\nu}(\tilde{\mathbf{r}})e^{\lambda_{\nu}t}$ and the corresponding phase fluctuation as $\delta S_{\nu}(\tilde{\mathbf{r}})e^{\lambda_{\nu}t}$, such that the constant $\lambda_{\nu}$ is an eigenvalue of $\mathcal{M}$. Given that expressing the action of the operator $\widehat{K}$ upon an arbitary monomial in $\mathbb{R}^3$ is somewhat cumbersome, we provide the relevant expression in Appendix~\ref{sec:level7}.

Since the time-dependence of the collective modes is exponential, the dynamical stability of a stationary state is determined to linear order by the set of all eigenvalues of $\mathcal{M}$, $\lbrace\lambda_{\nu}\rbrace$. If a given eigenvalue has a positive real component, the amplitude of the corresponding collective mode grows exponentially in time and overwhelms the stationary state, resulting in a dynamical instability. Therefore a stationary state is dynamically stable only if all of the eigenvalues of $\mathcal{M}$ have a negative real component, whereas purely imaginary eigenvalues are characteristic of excitations with an infinite lifetime. To diagonalize $\mathcal{M}$, we expand $\delta n$ and $\delta S$ as polynomials in $\mathbb{R}^3$~\cite{prl_87_19_190402_2001, prl_98_15_150401_2007, pra_82_5_053620_2010}. Since it is not possible to consider all possible collective modes, we truncate the polynomial expansion of the fluctuations such that the maximum allowed order of the polynomials is $N_{\text{max}} = 16$, which proves to be sufficient for characterizing the dynamical stability in the parameter regime that we explore. However, we note that even if no unstable modes are found when using the given value of the truncation parameter, $N_{\text{max}}$, it is not a guarantee of dynamical stability as a higher value of $N_{\text{max}}$ may admit unstable collective modes. Furthermore, it is conceivable that nonlinear effects, which are not accounted for in this linearized scheme, may destabilize collective modes that are stable at linear order in the fluctuations.

\begin{figure}[htbp]
\includegraphics[width=\linewidth]{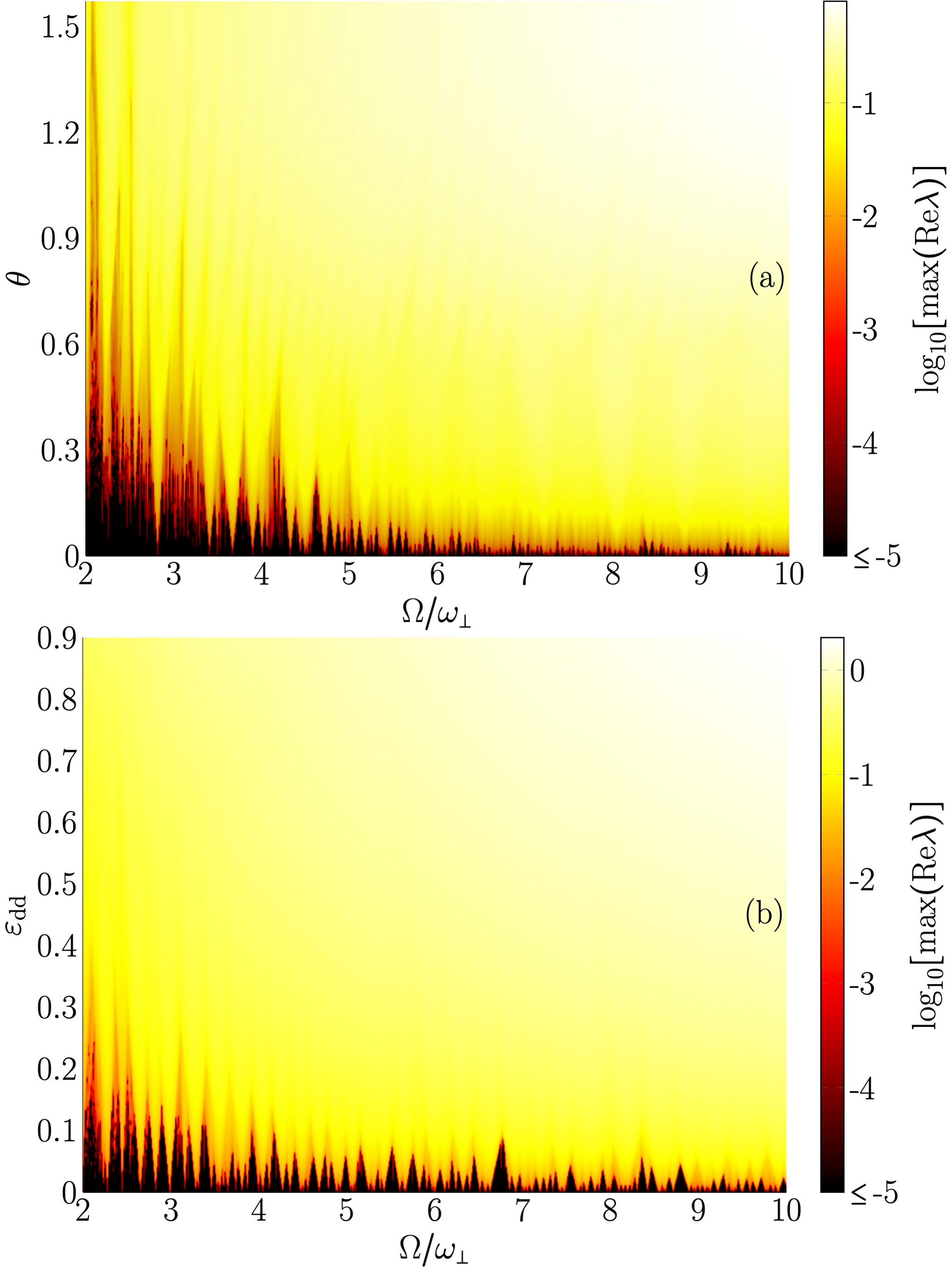}
\vspace*{-5mm}
\caption{Phase diagrams of the dynamical stability of the stationary solutions, with $N_{\text{max}} = 16$, for $\gamma = 1$ and $\Omega\in[2,10]\omega_{\perp}$. In (a), $\theta\in[0,\pi/2]$ and $\varepsilon_{\text{dd}} = 0.25$ while in (b), $\varepsilon_{\text{dd}}\in[0,0.9]$ and $\theta = \theta_{\text{c}}\equiv\arccos(1/\sqrt{3})\approx 0.955$. The condensate is dynamically unstable at all points in parameter space represented by a shade other than black. For increasing values of $\Omega$ and of $\theta$ (a), or $\varepsilon_{\text{dd}}$ (b), the largest positive real component of the eigenvalue spectrum is higher.}
\label{stability}
\end{figure}

We now proceed to diagonalize $\mathcal{M}$ with respect to the TF stationary solutions for a spherical trap ($\gamma = 1$) with $\Omega$ varied from $(1 + \gamma)\omega_{\perp} \equiv 2\omega_{\perp}$ to $10\omega_{\perp}$. With respect to $\varepsilon_{\text{dd}}$ and the dipole tilting angle, $\theta$, we choose to focus on two specific cases:
\begin{enumerate}
\item $\gamma = 1$, $\varepsilon_{\text{dd}} = 0.25$, $\Omega\in[2,10]\omega_{\perp}$, $\theta = [0, \pi/2]$, \label{en:constantedddiag}
\item $\gamma = 1$, $\theta = \theta_{\text{c}}$, $\Omega\in[2,10]\omega_{\perp}$, $\varepsilon_{\text{dd}} \in [0, 0.9]$. \label{en:constantthetadiag}
\end{enumerate}
Fixing $N_{\text{max}} = 16$, we have diagonalized Eq.~\eqref{eq:pertmatrix} and represented the results as phase diagrams of the dynamical stability of the corresponding stationary solutions in Figs.~\ref{stability} (a) and (b) for cases~\ref{en:constantedddiag} and \ref{en:constantthetadiag}, respectively. Specifically, Figs.~\ref{stability} (a) and (b) plot $\log_{10}\left[\max(\text{Re}\lambda)\right]$ as a function of $\widetilde{\Omega}$ and either $\theta$ (a) or $\varepsilon_{\text{dd}}$ (b). In both figures, if a given bin in parameter space is not shaded in black, the corresponding stationary solution is dynamically unstable against at least one collective mode and the condensate would be expected to eventually evolve away from the stationary state.

From Figs.~\ref{stability} (a) and (b), we can see that the stationary solution is unstable for all but the smallest values of either $\varepsilon_{\text{dd}}$ at constant $\theta = \theta_{\text{c}}$ or $\theta$ at constant $\varepsilon_{\text{dd}} = 0.25$. In both cases it is also evident that the domain of stability, at constant $\Omega$, becomes smaller for larger values of $\Omega$ and that the magnitude of the largest real, positive eigenvalues of $\mathcal{M}$ becomes larger as $\Omega$ increases. Furthermore, Figs.~\ref{stability} (a) and (b) demonstrate that higher values of $\theta$ and $\varepsilon_{\text{dd}}$, respectively, are characterized by unstable collective modes with larger real, positive eigenvalues. The existence of this dynamical instability is due to the ellipticity of the condensate about the rotation axis being nonzero for all $\theta > 0$ and $\varepsilon_{\text{dd}} > 0$~\cite{prl_122_5_050401_2019}. As a consequence of this ellipticity -- a combination of centrifugal distortion as well as magnetostriction -- the condensate undergoes solid-body rotation and is thus highly susceptible to processes such as the amplification of collective modes that lead to turbulence at rapid rotation frequencies. These results would suggest that it would be considerably difficult to achieve long-lived TF stationary states in this rotation frequency regime when $\theta \neq 0$ and/or $\varepsilon_{\text{dd}} > 0$. Although we do not conduct a parallel investigation of the numerical evolution of a dipolar BEC in this regime using the dGPE, such studies in the $\theta = \pi/2$ limit have demonstrated that the dipolar condensate evolves into a highly quantum turbulent state with an absence of clearly defined vortices after the dynamical instability has manifested itself~\cite{prl_122_5_050401_2019}. Therefore we expect that a similar scenario might occur when $\theta$ is nonzero.

We can also explore the nature of the dynamical instability by examining the expression given for the chemical potential in Eq.~\eqref{eq:chempotanswer} for a representative choice of the parameters, $\lbrace m, \varepsilon_{\text{dd}}, a_{\text{s}}, N, \omega_{\perp}\rbrace$. For the sake of reference we assume that $\varepsilon_{\text{dd}} = 0.75$ and use the parameters that correspond to the strongly dipolar species $^{162}$Dy~\cite{pra_92_2_022703_2015}, such that $a_{\text{s}} = 174.7\,a_0$, where $a_0$ is the Bohr radius. We also fix $N = 2\times10^4$ and $\omega_{\perp} = 2\pi\times 60$ Hz as suitably representative values for experimentally realized harmonically trapped BECs. Via Eqs.~\eqref{eq:chi}, \eqref{eq:chempotanswer}, \eqref{eq:kapparottuned}, and \eqref{eq:dipolarfunction} we find that the stationary state chemical potential for the time-averaged DDI is given by
\begin{align}
    \frac{\mu}{\hbar\omega_{\perp}} &= \frac{1}{2}\left[\frac{15\gamma Na_{\text{s}}}{(1-\overline{\varepsilon}_{\text{dd}})l_{\perp}}\right]^{\frac{2}{5}}\frac{\left\lbrace 1+\overline{\varepsilon}_{\text{dd}}[2-\frac{3}{2}(\gamma^2+2)F(\kappa)]\right\rbrace^{\frac{2}{5}}}{\left\lbrace 1+\overline{\varepsilon}_{\text{dd}}[2-3F(\kappa)]\right\rbrace^{\frac{3}{5}}} \nonumber \\
    &\times\left[1 - \overline{\varepsilon}_{\text{dd}}(1 - \kappa^2)F(\kappa)\right], \label{eq:chempottimeavg}
\end{align}
where $\kappa \equiv \kappa_{\llangle\text{DDI}\rrangle}$ is self-consistently determined by the solution of Eq.~\eqref{eq:kapparottuned}. For our choice of experimental parameters the chemical potential for the time-averaged DDI is therefore given by $\mu/(\hbar\omega_{\perp}) \approx 11.83$. This would suggest that experimental implementations of rotational tuning of the dipolar interaction would need to be conducted with dipole rotation frequencies at least $100$-$200$ times the in-plane trapping frequency in order for a meaningful number of rotation cycles to occur before a dynamical instability manifests itself. This is consistent with the experimental results of Tang, et al., in which a significantly reduced lifetime was reported for a condensate of $N = 2\times 10^4$ atoms with a dipole rotation frequency $\Omega = 2\pi\times 1000$ Hz, in comparison to trapping frequencies along $x$ and $y$ of $2\pi\times 73$ Hz and $2\pi\times 37$ Hz~\cite{prl_120_23_230401_2018}, respectively. This corresponds to $\widetilde{\Omega} = 1000/\sqrt{(73^2+37^2)/2} \approx 1000/57.87 \approx 17.28$, which is of the same order of magnitude as $\mu/(\hbar\omega_{\perp})$. Indeed, for larger condensates of $N = 10^5$ atoms, the TF chemical potential for the time-averaged DDI becomes $\mu/(\hbar\omega_{\perp}) \approx 22.51$, which necessitates even larger dipole rotation frequencies.

\section{\label{sec:level5}Conclusion}
In this work, we have extended the Thomas-Fermi theory for harmonically trapped dipolar Bose-Einstein condensates polarized by a continuously rotating applied field by allowing for the field to be oriented at any angle to its rotation axis. Inspired by analogous work on nondipolar BECs confined by a rotating, tilted, anisotropic harmonic trap~\cite{pra_101_6_063638_2020}, we find that the superfluid hydrodynamic equations describing this system may be solved exactly if we allow the condensate density's body frame to be tilted at an arbitary angle to the trapping axes. It is also necessary to allow for two, rather than one, degrees of freedom in the quadrupolar, irrotational velocity profile of the stationary state when the dipole rotation frequency is nonzero, and this results in a closed set of self-consistency relations that can be solved semi-analytically. In the limit of zero rotation, we recover the existing theory for the stationary states in the TF regime~\cite{pra_71_3_033618_2005, pra_82_3_033612_2010, pra_82_5_053620_2010}. For nonzero rotation frequencies, we have found that the self-consistency relations predict the existence of five distinct stationary solution branches as well as two distinct bifurcations of the solutions, rather than three branches and one bifurcation. The previously unknown bifurcation is located at a higher rotation frequency $\Omega_{\text{b}2}$ than the in-plane trapping frequency $\omega_{\perp}$ and is associated with an energetic instability of the $l = 2,\,m_z = 1$ surface mode, with the two previously unknown stationary solutions, branches III and IV, emerging from this new bifurcation. They exist for any dipole tilting angle and are always characterized by a nonzero tilt of the density profile; their existence when the dipole alignment is parallel or orthogonal to the rotation axis was previously hidden by the incomplete \textit{Ans{\"a}tze} employed in the literature. The remaining three branches reduce to the previously known stationary solutions for parallel or orthogonal dipole alignments in either limit~\cite{pra_80_3_033617_2009, pra_100_2_023625_2019}.

We have also found that when the rotation frequency is orders of magnitude larger than the in-plane harmonic trapping frequency, the stationary solutions are consistent with those obtained by considering non-rotating dipole moments subject to the time-averaged dipolar interaction. Crucially, this confirms that the stationary states mimic those of a nondipolar BEC when the dipole moments are aligned at an angle $\theta_{\text{c}} = \arccos{1/\sqrt{3}} \approx 54.7 \deg$ to the rotation axis. However in this ultrarapid rotational regime the stationary solutions are dynamically unstable against collective modes for all but the smallest values of $\varepsilon_{\text{dd}}$ and $\theta$. Hence, the condensate is unstable in the TF regime for ultrarapid dipolar rotation frequencies when $\theta = \theta_{\text{c}}$ even though the stationary solution approaches the nondipolar value. This instability is an artefact of the ellipticity of the density about the rotation axis that is induced by the tilting of the dipole moments. By calculating the TF value of the chemical potential for a BEC subject to the time-averaged dipolar interaction for a relevant choice of experimental parameters, we find that the chemical potential is generally $10$-$20$ times larger than $\hbar\omega_{\perp}$. Thus, it would be necessary for the dipole rotation frequency to be at least two orders of magnitude larger than the in-plane trapping frequency for a rotationally tuned TF stationary state to be stable for a substantial number of rotation cycles. We also expect these predictions to be reasonably robust when beyond mean-field quantum fluctuations of the ground state energy are accounted for; the resulting energy correction, which stabilises a dipolar condensate when $\varepsilon_{\text{dd}}$ and leads to the formation of ultradilute quantum droplets~\cite{nature_530_7589_194-197_2016, nature_539_7628_259-262_2016, prl_116_21_215301_2016, prx_6_4_041039_2016}, results in perturbative corrections to the collective modes~\cite{pra_86_6_063609_2012} and does not change their qualitative character.

\begin{acknowledgments}
S. B. P. is supported by the University of Melbourne's Melbourne Research Scholarship. The stability phase diagrams were determined via the Spartan HPC cluster, and we thank Research Computing Services at the University of Melbourne for access to this resource. S. B. P. thanks the research group of Benjamin Lev at the Departments of Physics and Applied Physics, Stanford University, for their hospitality and for several fruitful discussions during the initial stages of the research presented here, as well as the Laby Foundation for the financial support that enabled the visit. We also thank Nick Parker and Thomas Bland for the numerous stimulating discussions that motivated this work, as well as Mitchell Knight for providing feedback during the drafting of this paper.
\end{acknowledgments}

\appendix
\section{\label{sec:level6}Endpoints of the Stationary Solution Branches}
For four of the five stationary solution branches, we may define branch endpoints via the following:
\begin{enumerate}[label=(\alph*)]
\item $\tilde{\omega}_x \rightarrow 0$ and $\tilde{\omega}_y,\text{}\tilde{\omega}_z \neq 0$ $\Rightarrow\,\kappa_x\rightarrow\infty$, \label{en:endpointxapp}
\item $\tilde{\omega}_y \rightarrow 0$ and $\tilde{\omega}_x,\text{}\tilde{\omega}_z \neq 0$ $\Rightarrow\,\kappa_y\rightarrow\infty$, \label{en:endpointyapp}
\item $\tilde{\omega}_x,\text{}\tilde{\omega}_y \rightarrow 0$ and $\tilde{\omega}_z \neq 0$ $\Rightarrow\,\kappa_x,\,\kappa_y\rightarrow\infty$, \label{en:endpointxyapp}
\item $\tilde{\omega}_y,\text{}\tilde{\omega}_z \rightarrow 0$ and $\tilde{\omega}_x \neq 0$ $\Rightarrow\,\kappa_x\rightarrow 0$. \label{en:endpointyzapp}
\end{enumerate}
Let us denote the critical values of quantities such as $\widetilde{\Omega}$, $\tilde{\alpha}$, $\tilde{\delta}$, and $\xi$ at the limiting cases~\ref{en:endpointxapp}, \ref{en:endpointyapp}, \ref{en:endpointxyapp} and \ref{en:endpointyzapp} by the subscripts $xc$, $yc$, $xyc$, and $yzc$ respectively.

To characterize these critical values, it is necessary to calculate the respective values of $\kappa_x^p\kappa_y^q\beta_{ijk}:\,p,q,i,j,k\in\mathbb{Z}^{+}$ as they appear in Eqs.~\eqref{eq:zetax} -- \eqref{eq:zetaz} and \eqref{eq:xzcoeffzero}. Having evaluated these limits, we summarize them in Table~\ref{tab:kxkybeta}. Note that the necessity of having to evaluate the additional limit, $\tilde{\omega}_z^2 \rightarrow 0$ and $\tilde{\omega}_x,\text{}\tilde{\omega}_y \neq 0$, has been anticipated. In this limit both $\kappa_x$ and $\kappa_y\rightarrow 0$ and thus we have defined $K$ such that $\kappa_y\sim K\kappa_x$. Initially, we use Table~\ref{tab:kxkybeta} to check whether or not the right-hand side of Eq.~\eqref{eq:xzcoeffzero} is nonzero at a given endpoint for an arbitrary value of $\varepsilon_{\text{dd}}$. For case~\ref{en:endpointxapp}, $\kappa_x\kappa_y\beta_{101} \rightarrow 0$ and thus the right-hand side of Eq.~\eqref{eq:xzcoeffzero} is zero when $\tilde{\omega}_x^2 \rightarrow 0$. A similar identity holds for case~\ref{en:endpointxyapp} where, given that $\kappa_x\rightarrow\infty$ and $\kappa_y\rightarrow\infty$, $\kappa_x\kappa_y\beta_{101}\rightarrow 0$ in this limit. The right-hand side of Eq.~\eqref{eq:xzcoeffzero} is trivially zero for case~\ref{en:endpointyzapp} since $\tilde{\omega}_z^2 \rightarrow 0$. However, for case~\ref{en:endpointyapp}, we have $\kappa_x\kappa_y\beta_{101} \rightarrow 2/(1 + \kappa_x)^2$, suggesting that the properties of this limit are not universal with respect to $\varepsilon_{\text{dd}}$.

\begin{table*}
\caption{\label{tab:kxkybeta}Limits of $\kappa_x^p\kappa_y^q\beta_{ijk}:\,\lbrace p,q,i,j,k\rbrace\in\mathbb{Z}^{+}$, where the first and last rows correspond to the branch endpoint case~\ref{en:endpointxapp} and the second to fourth rows correspond to cases~\ref{en:endpointyapp} -- \ref{en:endpointyzapp}, respectively. These relations allow for the evaluation of the self-consistent evaluation of the limits of $\kappa_i$, $\tilde{\alpha}$, $\tilde{\delta}$, and $\xi$ at these endpoints.}
\begin{ruledtabular}
\begin{tabular}{c|cccccc}
& $\kappa_x^3\kappa_y\beta_{200}$ & $\kappa_x^3\kappa_y\beta_{101}$ & $\kappa_x\kappa_y^3\beta_{110}$ & $\kappa_x\kappa_y^3\beta_{011}$ & $\kappa_x\kappa_y\beta_{101}$ & $\kappa_x\kappa_y\beta_{002}$ \\
&&&\\[-1em] \hline
&&&\\[-0.75em]
$\kappa_x\rightarrow\infty$ & $0$ & $0$ & $0$ & $\dfrac{2\kappa_y^2}{(1+\kappa_y)^2}$ & $0$ & $\dfrac{2\kappa_y(2+\kappa_y)}{3(1+\kappa_y)^2}$ \\
&&&\\[-0.75em]
$\kappa_y\rightarrow\infty$ & $\dfrac{2(1+2\kappa_x)}{3(1+\kappa_x)^2}$ & $\dfrac{2\kappa_x^2}{(1+\kappa_x)^2}$ & $\dfrac{2}{1+\kappa_x}$ & $\dfrac{2\kappa_x}{1+\kappa_x}$ & $\dfrac{2}{(1+\kappa_x)^2}$ & $\dfrac{2\kappa_x(2+\kappa_x)}{3(1+\kappa_x)^2}$ \\
&&&\\[-0.75em]
$\kappa_x,\,\kappa_y\rightarrow\infty$ & $0$ & $2$ & $0$ & $2$ & $0$ & $\dfrac{2}{3}$ \\
&&&\\[-0.75em]
$\kappa_x\rightarrow 0$ & $\dfrac{2}{3}$ & $0$ & $2$ & $0$ & $2$ & $0$ \\
&&&\\[-0.75em]
$\kappa_x,\,\kappa_y\rightarrow 0$~\footnote{$\kappa_y\sim K\kappa_x$} & $\dfrac{2K(2+K)}{3(1+K)^2}$ & $0$ & $\dfrac{2K^2}{(1+K)^2}$ & $0$ & $\dfrac{2K}{1+K}$ & $0$ \\
\end{tabular}
\end{ruledtabular}
\end{table*}

For case~\ref{en:endpointxapp}, Eq.~\eqref{eq:alphadefn} implies that we have $\tilde{\alpha}_{xc} = \widetilde{\Omega}_{xc}\cos(\theta - \xi_{xc})$. Substituting this into Eqs.~\eqref{eq:omegaxeff} and \eqref{eq:xzcoeffzero} yields
\begin{align}
  (1-\widetilde{\Omega}_{xc}^2)\cos^2(\theta - \xi_{xc}) + \gamma^2\sin^2(\theta - \xi_{xc}) &= 0, \label{eq:omxzerolim1} \\
  (1-\gamma^2-\widetilde{\Omega}_{xc}^2)\sin[2(\theta - \xi_{xc})] &= 0. \label{eq:omxzerolim2}
\end{align}
Equations \eqref{eq:omxzerolim1} and \eqref{eq:omxzerolim2} admit the solution pair $\Omega_{xc} = \omega_{\perp}$ and $\sin(\theta-\xi_{xc}) = 0$, such that $\alpha_{xc} = \omega_{\perp}$ and $\delta_{xc} = 0$. We also expect the limit of $\kappa_y$ to be finite whereas $\kappa_x \rightarrow 0$. Using Eqs.~\eqref{eq:omegayeff} and \eqref{eq:omegazeff}, Eq.~\eqref{eq:kappaeqn} with $i \equiv y$, and the relevant limits in Table~\ref{tab:kxkybeta}, we find that the limit of $\kappa_y$ obeys the self-consistency relation
\begin{equation}
    \kappa_y^2 = \frac{\gamma^2\left[(1-\varepsilon_{\text{dd}})(1+\kappa_y)^2+3\varepsilon_{\text{dd}}\kappa_y^2\cos^2\theta\right]}{4\left[(1-\varepsilon_{\text{dd}})(1+\kappa_y)^2+3\varepsilon_{\text{dd}}\kappa_y(\kappa_y+2)\cos^2\theta\right]}. \label{eq:omxzerokyeqn}
\end{equation}

Since a rotation of the reference frame, $\tilde{\mathbf{r}}$, by $\pi/2$ about the $\tilde{y}$-axis would transform $R_x$ to $R_z$ and vice versa, we also solve for the condition that $\tilde{\omega}_z \rightarrow 0$ and $\tilde{\omega}_x,\text{}\tilde{\omega}_y \neq 0$, which must yield a limit that is physically identical to \ref{en:endpointxapp}. Let us denote the quantities in this limit by the subscript $zc$. From Eq.~\eqref{eq:deltadefn}, $\tilde{\delta}_{zc} = \widetilde{\Omega}_{zc}\sin(\theta-\xi_{zc})$. Substituting this into Eqs.~\eqref{eq:omegazeff} and \eqref{eq:xzcoeffzero} yields
\begin{align}
  (1-\widetilde{\Omega}_{zc}^2)\sin^2(\theta - \xi_{zc}) + \gamma^2\cos^2(\theta - \xi_{zc}) &= 0, \label{eq:omzzerolim1} \\
  (1-\gamma^2-\widetilde{\Omega}_{zc}^2)\sin[2(\theta - \xi_{zc})] &= 0. \label{eq:omzzerolim2}
\end{align}
Thus in this limit we have $\Omega_{zc} = \delta_{zc} = \omega_{\perp}$ and $\cos(\theta-\xi_{zc}) = \alpha_{zc} = 0$, which is consistent with the assumption that this limit is identical to that of \ref{en:endpointxapp} up to the rotation of the condensate density's body frame coordinates by $\pi/2$ about the $\tilde{y}$-axis. We also have $\kappa_x,\kappa_y\rightarrow 0$ in this limit and, by recognizing that $\sin^2(\theta-\xi_{zc}) = 1\,\Rightarrow\,\sin^2\xi = \cos^2\theta$, we can combine the relevant limits in Table~\ref{tab:kxkybeta} with Eqs.~\eqref{eq:omegaxeff}, \eqref{eq:omegayeff}, and \eqref{eq:kappaeqn} to obtain the self-consistency relation satisfied by $K$:
\begin{equation}
    K^2 = \frac{\gamma^2\left[(1-\varepsilon_{\text{dd}})(1+K)^2+3\varepsilon_{\text{dd}}K^2\cos^2\theta\right]}{4\left[(1-\varepsilon_{\text{dd}})(1+K)^2+3\varepsilon_{\text{dd}}K(K+2)\cos^2\theta\right]}. \label{eq:omzzerokykxeqn}
\end{equation}
As expected from the assumption that the limits $\tilde{\omega}_x\rightarrow 0$ and $\tilde{\omega}_z\rightarrow 0$ are physically equivalent, Equations~\eqref{eq:omxzerokyeqn} and\eqref{eq:omzzerokykxeqn} are identical.

For case~\ref{en:endpointyapp}, we find that we have
\begin{align}
  \alpha_{yc} &= -\Omega_{yc}\cos(\theta - \xi_{yc}), \label{eq:omyzeroaldel1} \\
  \delta_{yc} &= -\Omega_{yc}\sin(\theta - \xi_{yc}), \label{eq:omyzeroaldel2}
\end{align}
and by substituting these into Eq.~\eqref{eq:omegayeff}, we find that
\begin{equation}
  \Omega_{yc} = \omega_{\perp}. \label{eq:omyzerolimOm}
\end{equation}
Therefore, Eq.~\eqref{eq:xzcoeffzero} becomes:
\begin{equation}
  (4 - \gamma^2)\sin[2(\theta-\xi_{yc})] = \frac{3\tilde{\omega}_z^2\varepsilon_{\text{dd}}\kappa_x\kappa_y\beta_{101}\sin(2\xi)}{\zeta_z}. \label{eq:omyzerosxzlhssimp}
\end{equation}
From Eqs.~\eqref{eq:omegazeff}, \eqref{eq:omyzeroaldel1}, and \eqref{eq:omyzeroaldel2}, one also finds that
\begin{align}
    \tilde{\omega}_x^2 &\rightarrow 4\cos^2(\theta-\xi_{yc}) + \gamma^2\sin^2(\theta-\xi_{yc}), \label{eq:omyzerolimomx} \\
    \tilde{\omega}_z^2 &\rightarrow \gamma^2\cos^2(\theta-\xi_{yc}) + 4\sin^2(\theta-\xi_{yc}). \label{eq:omyzerolimomz}
\end{align}
Together with the relevant limits in Table~\ref{tab:kxkybeta}, Eqs.~\eqref{eq:omyzerolimomx} and \eqref{eq:omyzerolimomz} allow us to restate Eqs.~\eqref{eq:omyzerosxzlhssimp} and \eqref{eq:kappaeqn}, with $i \equiv x$, in terms of $\kappa_x$ and $\xi_{yc}$:
\begin{widetext}
\begin{gather}
  (4 - \gamma^2)\sin[2(\theta-\xi_{yc})] = \frac{3[\gamma^2\cos^2(\theta-\xi_{yc}) + 4\sin^2(\theta-\xi_{yc})]\varepsilon_{\text{dd}}\sin(2\xi)}{(1-\varepsilon_{\text{dd}})(1+\kappa_x)^2+3\varepsilon_{\text{dd}}[\sin^2\xi+\kappa_x(\kappa_x+2)\cos^2\xi]}, \label{eq:omyzerosxzfinal} \\
  \kappa_x^2 = \frac{[\gamma^2\cos^2(\theta-\xi_{yc}) + 4\sin^2(\theta-\xi_{yc})]\left\lbrace(1-\varepsilon_{\text{dd}})(1+\kappa_x)^2+3\varepsilon_{\text{dd}}[(1+2\kappa_x)\sin^2\xi+\kappa_x^2\cos^2\xi]\right\rbrace}{[4\cos^2(\theta-\xi_{yc}) + \gamma^2\sin^2(\theta-\xi_{yc})]\left\lbrace(1-\varepsilon_{\text{dd}})(1+\kappa_x)^2+3\varepsilon_{\text{dd}}[\sin^2\xi+\kappa_x(\kappa_x+2)\cos^2\xi]\right\rbrace}. \label{eq:omyzerokxfinal}
\end{gather}
\end{widetext}
By substituting these solutions of Eqs.~\eqref{eq:omyzerosxzfinal} and \eqref{eq:omyzerokxfinal} into Eqs.~\eqref{eq:omyzeroaldel1} and \eqref{eq:omyzeroaldel2}, one is able to characterize the stationary solutions in the limit $\tilde{\omega}_y^2 \rightarrow 0$.

The limits~\ref{en:endpointxyapp} and \ref{en:endpointyzapp} are somewhat more involved. In case~\ref{en:endpointxyapp} we have $\delta_{xyc} = -\Omega_{xyc}\sin(\theta-\xi_{xyc})$, but the limit of $\alpha_{xyc}$ is not as obvious and must be found by solving Eq.~\eqref{eq:xzcoeffzero}. This gives us
\begin{equation}
  \tilde{\alpha}_{xyc} = \frac{(1-\gamma^2+\widetilde{\Omega}_{xyc}^2)\cos(\theta-\xi_{xyc})}{2\widetilde{\Omega}_{xyc}}. \label{eq:omxyzeroal}
\end{equation}
Substituting these relations into Eqs.~\eqref{eq:omegaxeff} and \eqref{eq:omegayeff} results in the system of equations given by:
\begin{align}
&\left[(1-\gamma^2) + 2(1+\gamma^2)\widetilde{\Omega}_{xyc}^2-3\widetilde{\Omega}_{xyc}^4\right]\cos^2(\theta-\xi_{xyc}) \nonumber \\
&+ 4\gamma^2\widetilde{\Omega}_{xyc}^2\sin^2(\theta-\xi_{xyc}) = 0, \label{eq:omxyzerolimomxi1} \\
&(1-\gamma^2+\widetilde{\Omega}_{xyc}^2)(1-\gamma^2+5\widetilde{\Omega}_{xyc}^2)\cos^2(\theta-\xi_{xyc}) \nonumber \\
&+ 4\widetilde{\Omega}_{xyc}^2\left[1-\widetilde{\Omega}_{xyc}^2\sin^2(\theta-\xi_{xyc})\right] = 0. \label{eq:omxyzerolimomxi2}
\end{align}
Solving these simultaneously for $\Omega$ and $\xi$ yields the limiting values,
\begin{gather}
\Omega_{xyc} = (1 + \gamma)\omega_{\perp}, \label{eq:omxyzeroOmsol} \\
\cos^2(\theta-\xi_{xyc}) = \frac{\gamma}{2 + \gamma}. \label{eq:omxyzeroxisol}
\end{gather}
Therefore, from Eq.~\eqref{eq:omxyzeroal}, we have
\begin{equation}
\alpha_{xyc} = \omega_{\perp}\sqrt{\frac{\gamma}{2 + \gamma}}, \label{eq:omxyzeroalsol}
\end{equation}
and from the relation, $\delta_{xyc} = -\Omega_{xyc}\sin(\theta-\xi_{xyc})$, we have
\begin{equation}
\delta_{xyc} = -(1+\gamma)\omega_{\perp}\sqrt{\frac{2}{2 + \gamma}}. \label{eq:omxyzerodelsol}
\end{equation}
Although both $\kappa_x$ and $\kappa_y$ diverge in this limit, one may show that they obey the relation,
\begin{equation}
    \left(\frac{\kappa_y}{\kappa_x}\right)^2 \sim \frac{\gamma}{2 + \gamma}, \label{eq:omxyzerokykxsol}
\end{equation}
through the substitution of Eqs.~\eqref{eq:omxyzeroOmsol} -- \eqref{eq:omxyzeroalsol} into Eq.~\eqref{eq:alphadefn}.
However it is important to note that the expressions appearing in Eqs.~\eqref{eq:omxyzeroalsol} and \eqref{eq:omxyzerodelsol} are dependent on the branches of $\theta-\xi_{xyc}$ that are selected when solving Eq.~\eqref{eq:omxyzeroxisol}, since there are four equally valid choices that lie in the principal branch $[-\pi/2, \pi/2)$, viz. $\arccos\left(\sqrt{\frac{\gamma}{2 + \gamma}}\right)$, $\arccos\left(-\sqrt{\frac{\gamma}{2 + \gamma}}\right)$, $-\arccos\left(\sqrt{\frac{\gamma}{2 + \gamma}}\right)$, and $-\arccos\left(-\sqrt{\frac{\gamma}{2 + \gamma}}\right)$.  \\

In case~\ref{en:endpointyzapp}, we have $\alpha = -\Omega\cos(\theta-\xi)$ and from solving Eq.~\eqref{eq:xzcoeffzero} we also find that
\begin{equation}
\tilde{\delta}_{yzc} = \frac{(1-\gamma^2+\widetilde{\Omega}_{yzc}^2)\sin(\theta-\xi_{yzc})}{2\widetilde{\Omega}_{yzc}}. \label{eq:omyzzerodel}
\end{equation}
The substitution of these relations into Eqs.~\eqref{eq:omegayeff} and \eqref{eq:omegazeff} results in the following system of equations:
\begin{align}
&\left[1-\gamma^2 + 2(1+\gamma^2)\widetilde{\Omega}_{yzc}^2-3\widetilde{\Omega}_{yzc}^4\right]\sin^2(\theta-\xi_{yzc}) \nonumber \\
&+ 4\gamma^2\widetilde{\Omega}_{yzc}^2\cos^2(\theta-\xi_{yzc}) = 0, \label{eq:omyzzerolimomxi1} \\
&(1-\gamma^2+\widetilde{\Omega}_{yzc}^2)(1-\gamma^2)+5\widetilde{\Omega}_{yzc}^2)\sin^2(\theta-\xi_{yzc}) \nonumber \\
&+ 4\widetilde{\Omega}_{yzc}^2\left[1-\widetilde{\Omega}_{yzc}^2\cos^2(\theta-\xi_{yzc})\right] = 0. \label{eq:omyzzerolimomxi2}
\end{align}
As in case~\ref{en:endpointxyapp}, solving these equations for $\Omega$ and $\xi$ yields $\Omega_{yzc}$ and $\xi_{yzc}$, which we find to be given by
\begin{align}
\Omega_{yzc} &= (1 + \gamma)\omega_{\perp}, \label{eq:omyzzeroOmsol} \\
\cos^2(\theta-\xi_{yzc}) &= \frac{2}{2 + \gamma}. \label{eq:omyzzeroxisol}
\end{align}
These solutions are, of course, simply the same ones from Eqs.~\eqref{eq:omxyzeroOmsol} and \eqref{eq:omxyzeroxisol}, albeit subject to a rotation about the $\tilde{y}$-axis by $\pi/2$. They also yield limiting forms for $\tilde{\alpha}$ and $\tilde{\delta}$ of the form,
\begin{align}
\tilde{\alpha}_{yzc} &= -(1+\gamma)\omega_{\perp}\sqrt{\frac{2}{2 + \gamma}}. \label{eq:omyzzeroalsol} \\
\tilde{\delta}_{yzc} &= \omega_{\perp}\sqrt{\frac{\gamma}{2 + \gamma}}. \label{eq:omyzzerodelsol}
\end{align}
Finally, by substituting Eqs.~\eqref{eq:omyzzeroOmsol}, \eqref{eq:omyzzeroxisol}, and \eqref{eq:omyzzerodelsol} into Eq.~\eqref{eq:deltadefn}, we may show that
\begin{equation}
    \kappa_y \rightarrow \frac{\gamma}{2 + \gamma}, \label{eq:omyzzerokysol}
\end{equation}
whereas $\kappa_x \rightarrow 0$ since $\tilde{\omega}_z^2\rightarrow 0$.

\section{\label{sec:level7}Dipolar Contribution to the Collective Modes}
While it is relatively simple to calculate the transformation of a given monomial in $\mathbb{R}^3$ by the nondipolar components of $\mathcal{M}$, the evaluation of the dipolar contribution is quite involved. Nevertheless, the action of $\widehat{K}$ upon such a monomial can be computed via methods that were originally developed for the study of classical gravitationally-bound ellipsoidal fluids~\cite{apj_166_441-445_1971, pra_82_3_033612_2010, pra_82_5_053620_2010} and, in this section, we specify it explicitly. For a monomial $x^iy^jz^k$, with $\lbrace i, j, k\rbrace \in \mathbb{Z}$, we rewrite the exponents as
\begin{equation}
i = 2\lambda + \delta_{\lambda}\,,\, j = 2\mu + \delta_{\mu}\,,\, k = 2\nu + \delta_{\nu}, \label{eq:ijkdecomp}
\end{equation}
where $\lbrace\delta_{\lambda}, \delta_{\mu}, \delta_{\nu}\rbrace \in \lbrace 0, 1\rbrace$. The integral in $\mathcal{M}$ is then given by
\begin{widetext}
\begin{gather}
\int_{\Gamma_{\text{TF}}}\frac{x'^iy'^jz'^k\,\mathrm{d}^3r'}{4\pi\left|\mathbf{r}-\mathbf{r}'\right|} = \frac{R_x^iR_y^jR_z^ki!j!k!}{2^{2\sigma-1}}\sum_{p=0}^{\sigma}\sum_{q=0}^{\sigma-p}\sum_{r=0}^{\sigma-p-q}\frac{(-2)^{p+q+r}x^{2p+\delta_{\lambda}}y^{2q+\delta_{\mu}}z^{2r+\delta_{\nu}}\Lambda_{pqr}^{(i,j,k)}}{(2p)!(2q)!(2r)!(\sigma-p-q-r)!(2p\delta_{\lambda}+1)(2q\delta_{\mu}+1)(2r\delta_{\nu}+1)}, \label{eq:dipolekernelaction} \\
\Lambda_{pqr}^{(i,j,k)} = \sum_{l=0}^{\lambda}\sum_{m=0}^{\mu}\sum_{n=0}^{\nu}\frac{(-2)^{l+m+n}R_x^{2l+\delta_{\lambda}}R_y^{2m+\delta_{\mu}}R_z^{2n+\delta_{\nu}}M_{l+p+\delta_{\lambda},m+q+\delta_{\mu},n+r+\delta_{\nu}}}{(2l)!(2m)!(2n)!(\lambda-l)!(\mu-m)!(\nu-n)!(2l\delta_{\lambda}+1)(2m\delta_{\mu}+1)(2n\delta_{\nu}+1)}, \label{eq:lambdakernelcoefficient} \\
M_{lmn} = (2l-1)!!(2m-1)!!(2n-1)!!\frac{\kappa_x\kappa_y\beta_{lmn}}{2R_z^{2(l+m+n-1)}}, \label{eq:depolcoeff}
\end{gather}
\end{widetext}
with $\sigma = \lambda + \mu + \nu + 1$. From Eq.~\eqref{eq:koperator}, the dipolar contribution to $\mathcal{M}$ is obtained by acting upon the expression on the RHS of Eq.~\eqref{eq:dipolekernelaction} with the differential operator
\begin{equation}
    \left(\widehat{\mathbf{B}}\cdot\nabla\right)^2 \equiv \sin^2\xi\frac{\partial^2}{\partial\tilde{x}^2} + \cos^2\xi\frac{\partial^2}{\partial\tilde{z}^2} + \sin(2\xi)\frac{\partial^2}{\partial\tilde{x}\partial\tilde{z}}. \label{eq:directderiv}
\end{equation}
Please note that we integrate over the domain $\Gamma_{\text{TF}}$ and not over $\lbrace\tilde{\mathbf{r}}\in\mathbb{R}^3:n_{\text{TF}}(\tilde{\mathbf{r}}) < 0\,\cap\,n_{\text{TF}}(\tilde{\mathbf{r}}) + x^iy^jz^k \geq 0\rbrace$ as this would involve second-order effects in $x^iy^jz^k$~\cite{pra_82_3_033612_2010, pra_82_5_053620_2010}.

\bibliography{main.bbl}
\end{document}